\documentclass[aps,prd,preprintnumbers,groupedaddress,nofootinbib,amssymb,notitlepage,eqsecnum]{revtex4-1}
\usepackage{here}
\usepackage[dvipdfmx]{graphicx}
\usepackage{bm}
\usepackage{amsmath}
\usepackage{color}
\usepackage{amsfonts}
\usepackage[subnum]{cases}
\usepackage{hyperref}

\begin{document}
\preprint{WUCG-22-08}

\newcommand{\newc}{\newcommand}
\newc{\be}{\begin{equation}}
\newc{\ee}{\end{equation}}
\newc{\ba}{\begin{eqnarray}}
\newc{\ea}{\end{eqnarray}}
\newc{\bea}{\begin{eqnarray*}}
\newc{\eea}{\end{eqnarray*}}
\newc{\D}{\partial}
\newc{\ie}{{\it i.e.} }
\newc{\eg}{{\it e.g.} }
\newc{\etc}{{\it etc.} }
\newc{\etal}{{\it et al.}}
\newc{\Mpl}{M_{\rm Pl}}
\newcommand{\nn}{\nonumber}
\newc{\ra}{\rightarrow}
\newc{\lra}{\leftrightarrow}
\newc{\lsim}{\buildrel{<}\over{\sim}}
\newc{\gsim}{\buildrel{>}\over{\sim}}
\def\rd{\mathrm{d}}
\newcommand{\re}[1]{\textcolor{red}{#1}}
\newcommand{\ma}[1]{\textcolor{magenta}{#1}}
\newcommand{\mm}[1]{\textcolor{cyan}{[MM:~#1]}}

\title{
Symmetry restoration in the vicinity of neutron stars with a nonminimal coupling 
}

\author{
Masato Minamitsuji$^{1}$ and Shinji Tsujikawa$^{2}$}

\affiliation{
$^1$Centro de Astrof\'{\i}sica e Gravita\c c\~ao - CENTRA, Departamento de F\'{\i}sica, Instituto Superior T\'ecnico - IST, Universidade de Lisboa - UL, Avenida Rovisco Pais 1, 1049-001 Lisboa, Portugal\\
$^2$Department of Physics, Waseda University, 3-4-1 Okubo, Shinjuku, Tokyo 169-8555, Japan}

\date{\today}

\begin{abstract}
We propose a new model of scalarized neutron stars (NSs) realized by a self-interacting scalar field $\phi$ nonminimally coupled to the Ricci 
scalar $R$ of the form $F(\phi)R$. The scalar field has a self-interacting potential and sits at its vacuum expectation value $\phi_v$ far away from the source. Inside the NS, the dominance of a positive nonminimal coupling over a negative mass squared of the potential leads to 
a symmetry restoration with the central field value $\phi_c$ close to $0$. This allows the existence of scalarized NS solutions connecting $\phi_v$ with $\phi_c$ whose difference is significant, whereas the field is located in the vicinity of  $\phi=\phi_v$ for weak gravitational stars. The Arnowitt-Deser-Misner mass and radius of NSs as well as the gravitational force around the NS surface can receive sizable corrections from the scalar hair, while satisfying local gravity constraints in the Solar system.  Unlike the original scenario of spontaneous scalarization induced by a negative nonminimal coupling, the catastrophic instability of cosmological solutions can be avoided. We also study the cosmological dynamics from the inflationary epoch to today and show that the scalar field $\phi$ finally approaches the asymptotic value $\phi_v$ without spoiling a successful cosmological evolution. After $\phi$ starts to oscillate about the potential minimum, the same field can also be the source for cold dark matter.
\end{abstract}

\maketitle

\section{Introduction}
\label{sec1}

Having the detection of gravitational waves from binary systems composed of black holes (BHs) and/or neutron stars (NSs) \cite{LIGOScientific:2016aoc,LIGOScientific:2017vwq}, 
we are now ready for testing physics on strong gravitational backgrounds
in the strong field regime \cite{Berti:2015itd,Berti:2018cxi,Barack:2018yly}. 
General Relativity (GR) is currently recognized as 
a fundamental theory describing the 
gravitational interaction, but it is not yet clear how much 
extent to GR is trustable in the vicinity of extreme compact objects. 
There are some alternative theories of gravity like scalar-tensor 
theories \cite{Horndeski:1974wa,Fujii:2003pa,Deffayet:2011gz,
Kobayashi:2011nu,Charmousis:2011bf,Kase:2018aps,Kobayashi:2019hrl} 
in which a new degree of freedom like a scalar field 
could modify the gravitational interaction through couplings to 
curvature invariants. Since the accuracy of GR has been well 
confirmed in the weak-field regimes, modified gravitational theories 
have to be constructed to be consistent with local gravity constraints in the Solar system \cite{DeFelice:2010aj,Clifton:2011jh,Joyce:2014kja,Will:2014kxa,Koyama:2015vza,Heisenberg:2018vsk}.

In the presence of a scalar field $\phi$ nonminimally coupled with the Ricci scalar $R$ of the form $F(\phi)R$, it is known that a phenomenon called spontaneous scalarization can occur for static and spherically symmetric NSs \cite{Damour:1993hw}, 
while recovering the GR behavior in the weak-field backgrounds.
Spontaneous scalarization is an interesting phenomenon in that the large deviation from GR manifests itself on strong gravitational 
backgrounds \cite{Sotani:2004rq,Sotani:2014tua,Minamitsuji:2016hkk}.
In the presence of a scalar Gauss-Bonnet coupling, 
scalarization can occur for non-rotating and rotating 
BHs  \cite{Doneva:2017bvd,Silva:2017uqg,Doneva:2017bvd,Antoniou:2017acq,Antoniou:2017acq,Antoniou:2017acq,Minamitsuji:2018xde,Cunha:2019dwb,Dima:2020yac,Herdeiro:2020wei,Berti:2020kgk} as well as NSs \cite{Doneva:2017duq}.
Spontaneous scalarization can take place 
with a scalar-gauge coupling 
$\alpha(\phi)F_{\mu \nu}F^{\mu \nu}/4$ 
for charged BHs \cite{Herdeiro:2018wub,Fernandes:2019rez}
and charged stars \cite{Minamitsuji:2021vdb}.
While the extension of spontaneous scalarization of NSs to the 
vector-field sector has been considered in the 
literature \cite{Annulli:2019fzq,Ramazanoglu:2017xbl,Ramazanoglu:2019gbz,Kase:2020yhw,Minamitsuji:2020pak},
it has been argued that these models generically suffer from ghost or gradient instabilities \cite{Garcia-Saenz:2021uyv,Silva:2021jya,Demirboga:2021nrc}.

In the original model of Damour and Esposito-Farese based on the nonminimal coupling $F(\phi)R$ \cite{Damour:1993hw}, the necessary conditions for the occurrence of NS scalarization are given by $F_{,\phi}(0)=0$ and $F_{,\phi \phi}(0)>0$, where $F_{,\phi}={\rm d}F/{\rm d}\phi$ and 
$F_{,\phi \phi}={\rm d}^2F/{\rm d}\phi^2$.
In general, there is a nonvanishing scalar-field branch $\phi(r) \neq 0$ 
that depends on the radial distance $r$ besides a GR branch $\phi(r)=0$. 
The effective field mass squared around $\phi=0$ is given by 
$m_{\rm eff}^2(0)=-M_{\rm pl}^2 F_{,\phi \phi}(0)R_0/2$, 
where $M_{\rm pl}$ is the reduced Planck mass 
and $R_0$ is the Ricci scalar at $\phi=0$. 
In the weak-field backgrounds, the field 
can stay in the GR branch due to the smallness of $R_0$.  
Inside extreme compact objects like NSs, the 
negative mass squared $m_{\rm eff}^2(0)<0$ induced by 
large values of $R_0$ can trigger a tachyonic instability toward the 
nontrivial branch $\phi (r)\neq 0$.

The typical choice of nonminimal couplings consistent with 
the first condition $F_{,\phi}(0)=0$ is 
$F(\phi)={\rm e}^{-\beta \phi^2/(2\Mpl^2)}$, where $\beta$ 
is a constant. To realize the second condition $F_{,\phi \phi}(0)>0$, i.e., 
$m_{\rm eff}^2(0)<0$, we require that $\beta<0$.
The studies in Refs.~\cite{Harada:1998ge,Novak:1998rk,Silva:2014fca} 
have shown that spontaneous scalarization can occur for the nonminimal coupling 
in the range $\beta \le -4.35$, irrespective of the NS equation of state (EOS).
On the other hand, the binary pulsar measurements of 
an energy loss through the dipolar radiation 
have put the bound $\beta \ge -4.5$ \cite{Freire:2012mg,Shao:2017gwu}. 
Then, the coupling constant $\beta$ is constrained to be in a limited range.

If we apply the above nonminimally coupled theory to cosmology, it is known that the scalar field 
is subject to a tachyonic instability for negative values of $\beta$ required for the occurrence of spontaneous scalarization \cite{Damour:1992kf,Damour:1993id}. 
Around $\phi=0$, the effective field mass squared is 
estimated as $m_{\rm eff}^2 (0) \simeq \beta R_0/2$, 
so that $m_{\rm eff}^2 (0)<0$ for $\beta<0$ expect the 
radiation-dominated era (where $R_0=0$). 
During inflation in which the Hubble expansion rate $H$ is nearly 
constant, we have $m_{\rm eff}^2 (0) \simeq 6 \beta H^2$
and hence the negative coupling of order $\beta \simeq -5$
leads to the exponential growth of $\phi$. 
This spoils the success of the standard inflationary paradigm. 
We note that the initial field value at the onset of inflation cannot be 
tuned to 0 due to the presence of scalar-field perturbations $\delta \phi$. 
Indeed, the perturbations $\delta \phi$ relevant to the scales of 
observed CMB temperature anisotropies are exponentially amplified 
after the Hubble radius crossing during inflation. 
The scalar field also increases during matter and dark energy 
dominated epochs. Hence the GR solution $\phi=0$ is not 
a cosmological attractor and the Solar-system constraints would be easily violated. 
The similar instability of cosmological 
solutions is present for spontaneously scalarized BHs realized by a scalar Gauss-Bonnet coupling \cite{Anson:2019uto,Franchini:2019npi,Antoniou:2020nax}.

There have been several attempts to reconcile NS spontaneous 
scalarizations with cosmology.
One scenario is to take into account higher-order polynomial 
corrections (like ${\cal O}(\phi^4)$) to the nonminimal coupling 
function $F(\phi)$ \cite{Anderson:2016aoi}. 
There is also a scalarization scenario based on a disformal 
coupling between the scalar field $\phi$ and matter \cite{Silva:2019rle}. 
In this case, however, it was shown that the large disformal coupling 
required for the cosmological evolution toward $\phi=0$ works to 
suppress the occurrence of spontaneous scalarization.

The other scalarization scenario, which is called an ``asymmetron'' model \cite{Chen:2015zmx}, is to introduce a mass term of the scalar field
in the original model of Damour and Esposito-Farese, 
where the effective potential of the scalar field $\phi$ could 
have a global minimum. 
In this scenario, there is a nonvanishing global minimum and 
the scalar field moves toward this point due to tachyonic 
instability during inflation.
After the Universe enters the radiation-dominated epoch, the scalar field decouples from 
matter and the global minimum shifts back to the origin of the effective potential.
As the Universe expands further during the matter era,  
the Hubble parameter drops below the mass of $\phi$. 
Then the scalar field undergoes a damped oscillation, after which 
the cosmological evolution approaches that of GR. 
Hence GR is a cosmological attractor in the present Universe, 
while in local high-density regions spontaneous scalarization can occur as 
in the original Damour-Esposito-Farese model.
Moreover, the oscillating scalar field can be a candidate 
for cold dark matter (CDM).

There is also another possibility for introducing a coupling between $\phi$ and
the inflaton $\chi$ of the form $g^2 \phi^2 \chi^2/2$, 
where $g$ is a coupling constant~\cite{Anson:2019ebp}. 
Then the effective field mass squared $m_{\rm eff}^2 (\phi)$ can be largely 
positive during inflation, in which case $\phi$ decreases 
exponentially toward 0. After the end of the radiation-dominated era, 
the field $\phi$ starts to increase by the tachyonic mass. 
Provided that the suppression of $\phi$ during inflation 
occurs sufficiently, however, it is possible that today's 
value of $\phi$ is below the limit constrained by Solar-system experiments. 
Although the four-point coupling $g$ larger than 
the order $10^{-5}$ can lead to viable cosmological dynamics including 
the reheating epoch after inflation \cite{Nakarachinda:2022tjj}, 
the nonminimal coupling constant $\beta$ still needs to be 
in a limited negative range.

In this paper, we propose a new mechanism for NS scalarizations 
realized by the presence of a self-interacting potential of the form 
$V(\phi)=m^2 f_B^2 [1+\cos(\phi/f_B)]$
besides the nonminimal coupling ${\rm e}^{-\beta \phi^2/(2\Mpl^2)}R$, 
where $m$ and $f_B$ are constants with mass 
dimension\footnote{
In fact, our model does not correspond to ``spontaneous scalarization'' 
in the strict sense. The term ``spontaneous scalarization'' is typically 
used for phenomena where an excitation of the scalar field is realized 
as a continuous phase transition from a GR solution to the other 
nontrivial branch. This means that there should be both the GR and 
non-GR solutions with a nontrivial scalar field profile in a given theory.
In our model, the solution approaching $\phi=\phi_v\,(=\pi f_B)$ 
at asymptotic infinity is not connected to a GR solution
with a continuous phase transition (see Fig.~\ref{fig5}). 
Nevertheless, in the whole manuscript, we call our solution ``the scalarized solution''
in the sense that it could overcome the difficulty 
of embedding the Damour--Esposito-Farese model 
into the realistic cosmic expansion history.
}.
In this setup, the field $\phi$ is in a ground state
at the vacuum expectation value (VEV)
$\phi_v=\pi f_B$ in the asymptotic region far away from a NS.
At $\phi=0$ the bare potential $V(\phi)$ has a negative 
mass squared $-m^2$, but the positive nonminimal coupling constant 
($\beta>0$) gives rise to a positive contribution $\beta R_0/2$ to 
the effective mass squared as $m_{\rm eff}^2(0)=-m^2+\beta R_0/2$. 
In the high-curvature region with $m_{\rm eff}^2(0)>0$, 
the field $\phi$ can stay in the vicinity of $\phi=0$. 
The transition to the region close to $\phi=0$ should occur 
inside the NS for the coupling $\beta>{\cal O}(0.1)$ 
with $m={\cal O}(10^{-11}\,{\rm eV})$, where the Compton 
radius $m^{-1}={\cal O} (10\,{\rm km})$ corresponds to 
the typical size of NSs.
We will show the existence of field profiles connecting the internal 
solution ($\phi \simeq 0$) to the external solution far outside 
the star ($\phi \simeq \phi_v$).  
A conceptually similar model was proposed in Ref.~\cite{Babichev:2022djd},
where scalarized BHs were induced by the scalar Gauss-Bonnet 
coupling with a symmetry-breaking potential.

We note that the structure of our model is similar to the symmetron 
scenario \cite{Hinterbichler:2010es}, which was proposed as 
one of the screening mechanisms of fifth forces 
in local regions of the Universe. 
The similarity is that positive nonminimal couplings are used to 
restore the symmetry at $\phi=0$ at high density and that 
the tachyonic mass of the potential breaks the symmetry 
to reach the state at $\phi=\phi_v$ at low density.
In the symmetron model the scalar field is relevant 
to the late-time cosmic acceleration (i.e., dark energy), 
so that the scalar field mass is as small as $m={\cal O}(10^{-30}~{\rm eV})$. 
The fifth force can be suppressed for $\phi$ close to 0, 
but it propagates once $\phi$ reaches the region close to $\phi_v$ 
(see Refs.~\cite{Burrage:2017qrf,Brax:2021wcv} for laboratory tests 
of the symmetron). 
In our model the typical mass scale is as large as 
$m={\cal O}(10^{-11}~{\rm eV})$, 
in which case it is possible to satisfy local gravity constraints 
even for $\phi$ close to $\phi_v$. 
In the vicinity of NSs, the scalar field can reach the region close to $\phi=0$ 
and hence the spherically symmetric solutions in strong gravity regimes 
exhibit differences from those in weak gravity regimes.
Cosmologically, the scalar field can also  
behave as CDM after the symmetry breaking.
These properties are different from those in the symmetron model.
Our model is also different from the asymmetron model mentioned above
in that the Universe approaches the GR 
vacuum at $\phi=0$ but not the one at $\phi=\phi_v$ 
in late-time/low density regimes.

In our model, the nonminimal coupling constant $\beta$ is positive 
and the effective mass squared $m_{\rm eff}^2(0)=-m^2+\beta R_0/2$ 
at $\phi=0$ is positive in the early cosmological epoch 
satisfying $\beta R_0/2>m^2$. 
Then, during inflation, the scalar field $\phi$ can decrease exponentially toward 0. After $\beta R_0/2$ drops below $m^2$ in the radiation-dominated era, the field $\phi$ should exhibit tachyonic growth toward 
the ground state at $\phi=\phi_v$. 
Indeed, we will show that the field settles down 
the potential minimum by today without violating a successful 
cosmic expansion history. After $\phi$ starts to oscillate around 
$\phi_v$, the same field can also 
work as the source for (a portion of) CDM. 

In weak gravitational objects like the Sun, the Ricci scalar $R$ 
inside the star is small in comparison to that in NSs 
and hence $m_{\rm eff}^2$ is negative in the vicinity 
of $\phi=0$. In such cases, the scalar field is 
in the region close to $\phi=\phi_v$ both inside and 
outside the star. We will obtain the field profile and post-Newtonian 
parameter and put bounds on the scale $f_B$ 
from Solar-system tests on local gravity.
Using these constrained values of $f_B$, we numerically 
construct scalarized NS solutions with nontrivial profiles of the scalar field and compute the effect on 
the Arnowitt-Deser-Misner (ADM) mass and radius of NSs as well as modifications of gravity around the surface of star.
We will show that the difference of the ADM mass of scalarized NSs from that in GR can exceed more than 10\,\%. 
The modified gravitational interaction induced in our scenario may be detectable in future observations of gravitational waves 
and other measurements on the strong field regimes.

This paper is organized as follows.
In Sec.~\ref{modelsec}, we present our new model of NS 
scalarizations and discuss its basic structure.
In Sec.~\ref{cossec}, we study the cosmological dynamics of 
the nonminimally coupled scalar field from an inflationary epoch to today 
and show that the field is eventually stabilized at $\phi=\phi_v$ 
without preventing the cosmic expansion history.
In Sec.~\ref{weaksec}, we derive the field profile for a constant 
density star on weak gravitational backgrounds and place 
bounds on model parameters from Solar-system constraints.
In Sec.~\ref{scasec}, we obtain the scalar-field solution
for NSs and study its effect on the modification of gravitational 
interactions. Sec.~\ref{consec} is devoted to conclusions.

\section{Models with NS scalarizations}
\label{modelsec}

We consider theories given by the action 
\ba
{\cal S}=
\int {\rm d}^4 x \sqrt{-g_J} 
\left[ \frac{\Mpl^2}{2}F(\phi) R+\omega (\phi)X
-V(\phi) \right]
+{\cal S}_m (g_{\mu \nu}, \Psi_m)\,,
\label{action}
\ea
where $g_J$ is a determinant of the metric tensor $g_{\mu \nu}$, 
$M_{\rm pl}$ is a constant having the dimension of mass, 
$F$ is a function of $\phi$, $R$ is the Ricci scalar, 
$X=-(1/2)g^{\mu \nu} \nabla_{\mu} \phi \nabla_{\nu} \phi$ 
is a scalar kinetic term with the covariant derivative operator 
$\nabla_{\mu}$, $V(\phi)$ is a scalar potential, and 
\be
\omega(\phi)=\left( 1-\frac{3\Mpl^2 F_{,\phi}^2}{2F^2} 
\right)F\,,
\label{omega}
\ee
with $F_{,\phi}\equiv {\rm d} F/{\rm d}\phi$ and so on.
The action ${\cal S}_m$ incorporates the contributions of 
matter fields $\Psi_m$ inside the NS.
Note that in the case $F(\phi)=1$ the constant $\Mpl$ 
represents the reduced Planck mass ($\Mpl=2.435\times 10^{18}$ GeV).

The equations of motion for the metric and scalar field 
are given, respectively, by
\ba
&&
\label{metric_eom}
\Mpl^2
\left[
F(\phi)G_{\mu\nu}
+\Box F (\phi)g_{\mu\nu}
-\nabla_\mu \nabla_\nu F(\phi)
\right]
-\omega(\phi)
\left(
\nabla_\mu\phi\nabla_\nu \phi
+X g_{\mu\nu}
\right)
+
g_{\mu\nu}V(\phi)
=
T_{\mu\nu},
\\
&&
\label{scalar_eom}
\omega(\phi)\Box\phi
-\omega_{,\phi} (\phi)X
+\frac{\Mpl^2}{2}F_{,\phi}(\phi)R
-V_{,\phi}(\phi)
=0,
\ea
where $T_{\mu\nu}$ represents the energy-momentum tensor 
of matter in the Jordan frame defined by 
\be
T_{\mu\nu}
\equiv -\frac{2}{\sqrt{-g_J}}
\frac{\delta {\cal S}_m}{\delta g^{\mu\nu}}.
\ee
Acting the operator $\nabla^\mu$ on Eq. \eqref{metric_eom} and 
using Eq.~\eqref{scalar_eom}, we obtain the conservation law of matter as
\be
\label{eq_continuity}
\nabla^\mu T_{\mu\nu}=0.
\ee

We consider the nonminimal coupling  
chosen by Damour and Esposito-Farese \cite{Damour:1993hw}
\be
F(\phi)={\rm e}^{-\beta \phi^2/(2M_{\rm pl}^2)}\,,
\label{Fnon}
\ee
where $\beta$ is a dimensionless constant. 
{}From Eq.~\eqref{omega}, we have 
\be
\omega(\phi)=\left( 1- \frac{3 \beta^2 \phi^2}{2 \Mpl^2} 
\right)F\,.
\ee
Under a conformal transformation of the action (\ref{action}) 
to the Einstein frame, the theory with $V(\phi)=0$ recasts the 
one originally advocated in Ref.~\cite{Damour:1993hw} 
(see the Appendix of Ref.~\cite{Kase:2020yhw}).
In the following, we will perform all the analysis in the Jordan 
frame action (\ref{action}).

Let us first revisit the case of standard NS spontaneous scalarization 
in the absence of the scalar potential, i.e., 
\be
V(\phi)=0\,.
\ee
Then, there is the branch $\phi=0$ as one of the solutions to 
Eq.~(\ref{scalar_eom}). For this solution, Eq.~(\ref{metric_eom}) 
reduces to the Einstein equation $\Mpl^2 G_{\mu \nu ({\rm GR})}=T_{\mu \nu ({\rm GR})}$ 
in GR. In regions of the large curvature $R$, it is possible to 
have a nontrivial branch with $\phi \neq 0$ besides the GR branch $\phi=0$.
If we consider a small perturbation $\delta \phi$ about the GR solution, 
the perturbation obeys $\square \delta \phi -m_{\rm eff}^2 (0) \delta \phi=0$, where 
$m_{\rm eff}^2(0)=-M_{\rm pl}^2 F_{,\phi \phi}(0)R_0/2$ 
and $R_0$ is the Ricci scalar in the GR background at $\phi=0$.
Provided that $F_{,\phi \phi}(0)>0$ with $R_0>0$, 
the GR branch is subject to tachyonic instability due to 
the negative mass squared $m_{\rm eff}^2(0)$. 
For $\beta<0$, there is a possibility for NSs to acquire a scalar hair 
after the spontaneous growth of $\phi$ toward the other nontrivial branch, 
whose phenomenon is dubbed spontaneous scalarization.

As mentioned in Sec.~\ref{sec1}, the nonminimal coupling constant $\beta$ 
needs to be in the limited range $-4.5 \le \beta \le -4.35$ in the original 
model of Ref.~\cite{Damour:1993hw}. Here, the upper limit of $\beta$ 
arises for the occurrence of spontaneous scalarization \cite{Harada:1998ge,Novak:1998rk}, whereas the lower bound comes from binary pulsar measurements \cite{Freire:2012mg,Shao:2017gwu}. 
For such negative values of $\beta$, there is a tachyonic instability of the field 
$\phi$ on the cosmological background and hence GR solution $\phi=0$ is not an attractor. This instability is particularly prominent during the inflationary epoch to destroy  the background evolution. Then, the successful cosmic expansion history is spoiled by the negative 
nonminimal coupling with $V(\phi)=0$.

The story is different for the positive nonminimal coupling 
with a self-interacting scalar potential $V(\phi)$. 
For concreteness, we consider a potential of 
the pseudo Nambu-Goldstone boson (pNGB), 
which is given by 
\be
V(\phi)=m^2 f_B^2 \left[ 1+\cos \left( \frac{\phi}{f_B} 
\right) \right]\,,
\label{Vphi}
\ee
where $m$ and $f_B$ are constants having the dimension of mass. 
This potential has a reflection symmetry with respect to $\phi=0$. 
To choose either the ground state at $\phi=\pi f_B$ or $\phi=-\pi f_B$ 
means the breaking of the reflection symmetry. 
We will choose the positive VEV $\phi_v=\pi f_B$ as a symmetry-breaking 
ground state.
Note that in order to test our idea,
the form of the scalar field potential 
may not be restricted to a particular from as Eq.~\eqref{Vphi}.
We can consider other symmetry-breaking potentials like
$V(\phi)=h^2(\phi^2-\phi_v^2)^2$, 
where $h$ and $\phi_v$ are constants. 
Indeed, so long as the potential has a local maximum at $\phi=0$
and a global minimum at $\phi\neq 0$, it is sufficient for our purpose of 
the construction of a viable model.
We choose the particular pNGB potential \eqref{Vphi} for an illustration.

Around $\phi=0$, the potential (\ref{Vphi}) has a negative mass squared $-m^2$. Since the nonmininal coupling $\Mpl^2 F(\phi)R/2$ is present, the squared effective mass of the field at $\phi=0$ yields
\be
\label{m0}
m_{\rm eff}^2(0)=-m^2+\frac{\beta}{2}R_0\,.
\ee
Due to the largeness of $R_0$ in regions of the high density, 
the positive nonminimal coupling constant $\beta$ can lead to 
the symmetry restoration at $\phi=0$. 
This occurs if $\beta R_0/2$ exceeds the negative mass 
squared $-m^2$. In regions of the low density, the effect of 
$\beta R_0/2$ on $m_{\rm eff}^2(0)$ should be unimportant relative to 
the contribution $-m^2$. 
Hence the scalar field would acquire the VEV $\phi_v=\pi f_B$ on 
weak gravitational backgrounds.
This scalar-field configuration is different from that arising 
from standard NS spontaneous scalarization with $V(\phi)=0$, 
in that the scalar field is in the symmetry-restored state $\phi=0$ around the center of
star
while $\phi$ approaches the asymptotic value 
$\phi_v=\pi f_B$ far away from the star. 

For a star with the mean density $\rho$ and pressure $P$, 
the Ricci scalar $R$ at $\phi=0$ is of order $R\simeq (\rho-3P)/\Mpl^2$. 
Then, the critical value of $\beta$ corresponding to $m_{\rm eff}^2(0)=0$ can be estimated as
\be
\beta_c=\frac{2 m^2 \Mpl^2}{\rho-3P}
=0.28
\left( \frac{10^{15}~{\rm g/cm}^3}{\rho-3P} \right) 
\left( \frac{m}{10^{-11}~{\rm eV}} \right)^2\,.
\label{betac}
\ee
Note that, for $m={\cal O}(10^{-11}~{\rm eV})$, the Compton radius of 
$\phi$ is of ${\cal O}(10\,{\rm km})$, i.e., the typical the size of NSs. 
For $\beta>\beta_c$ we have $m_{\rm eff}^2(0)>0$, and
the scalar field can be in the symmetry-restored state at $\phi=0$. 
For $\beta<\beta_c$, the state at $\phi=0$ becomes unstable 
and hence the solution should approach the ground state at 
$\phi=\phi_v$.
The typical central density of NSs is around $\rho=10^{15}~{\rm g/cm}^3$, 
so the mass of order $m=10^{-11}~{\rm eV}$ gives rise to the critical 
coupling $\beta_c$ around $\beta_c=0.1\sim 1$.

On the Friedmann-Lema\^itre-Robertson-Walker (FLRW) cosmological 
background, the scalar field can be 
in the state $\phi=0$ in the early Universe satisfying the condition 
$\beta R_0/2>m^2$. After the term $\beta R_0/2$ drops below $m^2$ 
along the cosmic expansion, however, the field should evolve 
to the ground state at $\phi=\phi_v$ since 
$m_{\rm eff}^2(0)$ becomes negative. 
In Sec.~\ref{cossec}, we study cosmology in 
the above model in details and show that $\phi$ sufficiently 
approaches the potential minimum by today.

\section{Cosmology with positive nonminimal coupling}
\label{cossec}

We study the cosmological dynamics of the scalar field $\phi$ from 
the inflationary epoch to today for the theory given by the action (\ref{action}). 
A spatially-flat FLRW background is 
given by the line element 
\be
{\rm d}s^2
=g_{\mu\nu}{\rm d}x^\mu {\rm d}x^\nu
=-{\rm d}t^2+a^2(t) \delta_{ij}
{\rm d}x^i {\rm d}x^j\,,
\ee
where the scale factor $a(t)$ depends on the cosmic time $t$.
Then, the gravitational and scalar-field equations of motion are 
\ba
& & 3F H^2 \Mpl^2=
-3\Mpl^2 H F_{,\phi}\dot{\phi}
+\frac{1}{2} \omega \dot{\phi}^2+V+\rho\,,
\label{cosmo1}\\
& &
F \left( 2\dot{H}+3H^2 \right) \Mpl^2=
-\Mpl^2
\left[
F_{,\phi} (\ddot{\phi}+2H\dot{\phi})
+F_{,\phi\phi}\dot{\phi}^2
\right]
-\frac{1}{2}\omega \dot{\phi}^2+V-P\,,
\label{cosmo2}\\
& &
\ddot{\phi}+3H \dot{\phi}
-\frac{3\Mpl^2F_{,\phi}}{\omega}
\left(\dot{H}+2H^2\right)
+\frac{\omega_{,\phi}\dot{\phi}^2}{2\omega}
+\frac{V_{,\phi}}{\omega}=0\,,
\label{cosmo3}
\ea
where 
$\rho$ and $P$ are the density and pressure of the 
inflaton field and/or perfect fluids, $H=\dot{a}/a$ 
is the Hubble parameter, and a `dot' represents the 
derivative with respect to $t$, and 
$\omega_{,\phi}\equiv {\rm d}\omega/{\rm d}\phi$, 
$V_{,\phi}\equiv {\rm d} V/{\rm d}\phi$, and so on.
Note that
$F_{,\phi}= -\beta \phi F/\Mpl^2$ and 
$F_{,\phi\phi}=\beta (\beta \phi^2-\Mpl^2)F/\Mpl^4$,
and in the regime $|\phi|\ll \Mpl$,
$\omega_{,\phi}/\omega \simeq -\beta(1+3\beta)\phi/\Mpl^2$.

\subsection{Evolution during inflation and reheating}

To study the cosmological dynamics during inflation, we incorporate 
a canonical inflaton field $\chi$ with the potential $U(\chi)$.  
Then, we have $\rho=\dot{\chi}^2/2+U(\chi)$ and 
$P=\dot{\chi}^2/2-U(\chi)$ in Eqs.~(\ref{cosmo1}) and (\ref{cosmo2}). 
The inflaton field obeys the continuity equation $\dot{\rho}+3H (\rho+P)=0$, i.e., 
\be
\ddot{\chi}+3H\dot{\chi}+U_{,\chi}=0\,,
\label{chieq}
\ee
where $U_{,\chi}\equiv {\rm d} U/{\rm d}\chi$.
The kinetic and potential energy of the field $\phi$ 
should be suppressed relative to $U(\chi)$ during inflation.
Let us consider the typical Hubble scale of inflation 
of order $H \sim 10^{14}$~GeV. 
Since $V(\phi)$ is at most of order $m^2 f_{B}^2$, 
we have $V(\phi) \lesssim m^2 f_B^2 \ll H^2 \Mpl^2 \sim U(\chi)$ 
for the mass scale $m \sim 10^{-11}$~eV with 
$f_B \lesssim \Mpl$. Provided that the condition 
$|\omega_{,\phi}|\dot{\phi}^2 \ll H^2\Mpl^2|F_{,\phi}|$ holds 
together with the slow-roll condition $|\dot{H}| \ll H^2$, 
Eq.~(\ref{cosmo3}) is approximately given by 
\be
\ddot{\phi}+3H \dot{\phi}+\frac{1}{\omega} 
\left[ 6 \beta F H^2 \phi -m^2 f_B \sin \left( \frac{\phi}{f_B} 
\right)\right] \simeq 0\,.
\label{phiinf}
\ee
We are interested in the coupling range $\beta \gtrsim 0.1$ 
with $m$ of order $10^{-11}$~eV. 
For $\phi \gtrsim f_B$, since $H \gg m$, the term $6 \beta F H^2 \phi$ 
dominates over $m^2 f_B \sin (\phi/f_B)$ during inflation.
This is also the case for $0<\phi \ll f_B$ as 
$m^2 f_B \sin (\phi/f_B) \simeq m^2 \phi$ in this regime.
Then, during inflation, Eq.~(\ref{phiinf}) approximately reduces to 
\be
\ddot{\phi}+3H \dot{\phi}+\frac{6 \beta F H^2}{\omega} 
\phi \simeq 0\,,
\label{phiin}
\ee
and the contribution of the pNGB scalar potential to 
the background Eqs.~(\ref{cosmo1})-(\ref{cosmo2}) can be 
completely neglected. 
Provided that the scalar field is in the range $\beta \phi^2/\Mpl^2 \ll 1$, 
we have $F \simeq 1$ and $\omega \simeq 1$. 
On using the approximation that $H$ is constant during inflation, 
the dominant solution to Eq.~(\ref{phiin}) is given by 
\begin{numcases}{\phi \propto }
     \exp \left( -\frac{3}{2} H t \right) \cos(\Omega_0 t+\theta_0)    &   
     ({\rm if}~$\beta>3/8)$\,, \label{dampos} \\
     \exp \left[ -\frac{3}{2} \left( 1-\sqrt{1-\frac{8}{3}\beta} 
\right)H t \right]   &   ({\rm if}~$\beta < 3/8)$\,,\label{phidec}
\end{numcases}
where $\Omega_0=\sqrt{6\beta-9/4}\,H$ and $\theta_0$ is an arbitrary constant. 
For $\beta>3/8$, the field $\phi$ exhibits a damped oscillation 
with the amplitude rapidly decreasing as $|\phi| \propto \exp (-3H t/2)$. 
If the total number of e-foldings during inflation is
$N=\int_0^t H{\rm d}t \simeq Ht=60$, the amplitude of $\phi$
at the end of inflation is $8 \times 10^{-40}$ times as small 
as that at the onset of inflation. 
For $0<\beta<3/8$, $\phi$ decreases without oscillations 
according to Eq.~(\ref{phidec}).
If $\beta<0$, the scalar field increases 
as $\phi \propto \exp[(3/2) (\sqrt{1-8\beta/3}-1)Ht]$.
 
As the inflaton potential, we consider the $\alpha$-attractor type 
given by  
\be
U(\chi)=\frac{3}{4}\alpha M^2 \Mpl^2 
\left[ 1-\exp \left( -\sqrt{\frac{2}{3\alpha}} \frac{\chi}{\Mpl} 
\right) \right]^2\,,
\label{Vchi}
\ee
where $\alpha$ is a positive constant \cite{Kallosh:2013yoa}. 
For $\alpha=1$, the potential (\ref{Vchi}) is equivalent to 
that of Starobinky inflation \cite{Starobinsky:1980te} 
in the Einstein frame \cite{DeFelice:2010aj}.
The field $\chi$ at the end of inflation can be
determined by the condition $\epsilon_V=(\Mpl^2/2)(V_{,\chi}/V)^2=1$, 
i.e., $\chi_f=0.940 \Mpl$. Cosmic acceleration occurs in the region 
where $\chi \gtrsim \Mpl$, which is followed by the reheating stage 
driven by the oscillation of $\chi$ around $\chi=0$.
{}From the Planck normalization of curvature perturbations generated 
during inflation, the mass $M$ is constrained to be around $M \simeq 10^{-5} \Mpl$. 
In our numerical simulations we will choose the potential (\ref{Vchi}) 
with $\alpha=1$, but the evolution of $\phi$ during inflation and reheating 
is similar for any other slow-roll inflaton potentials which can be approximated by 
$U(\chi) \simeq M^2 \chi^2/2$ in the vicinity of $\chi= 0$. 
Indeed, the analytic solutions to $\phi$ given by Eqs.~(\ref{dampos}) and (\ref{phidec})
during inflation as well as those during reheating derived later in 
Eqs.~(\ref{damposre}) and (\ref{phidecre}) are insensitive 
to the change of inflaton potentials.

\begin{figure}[ht]
\begin{center}
\includegraphics[height=3.2in,width=3.4in]{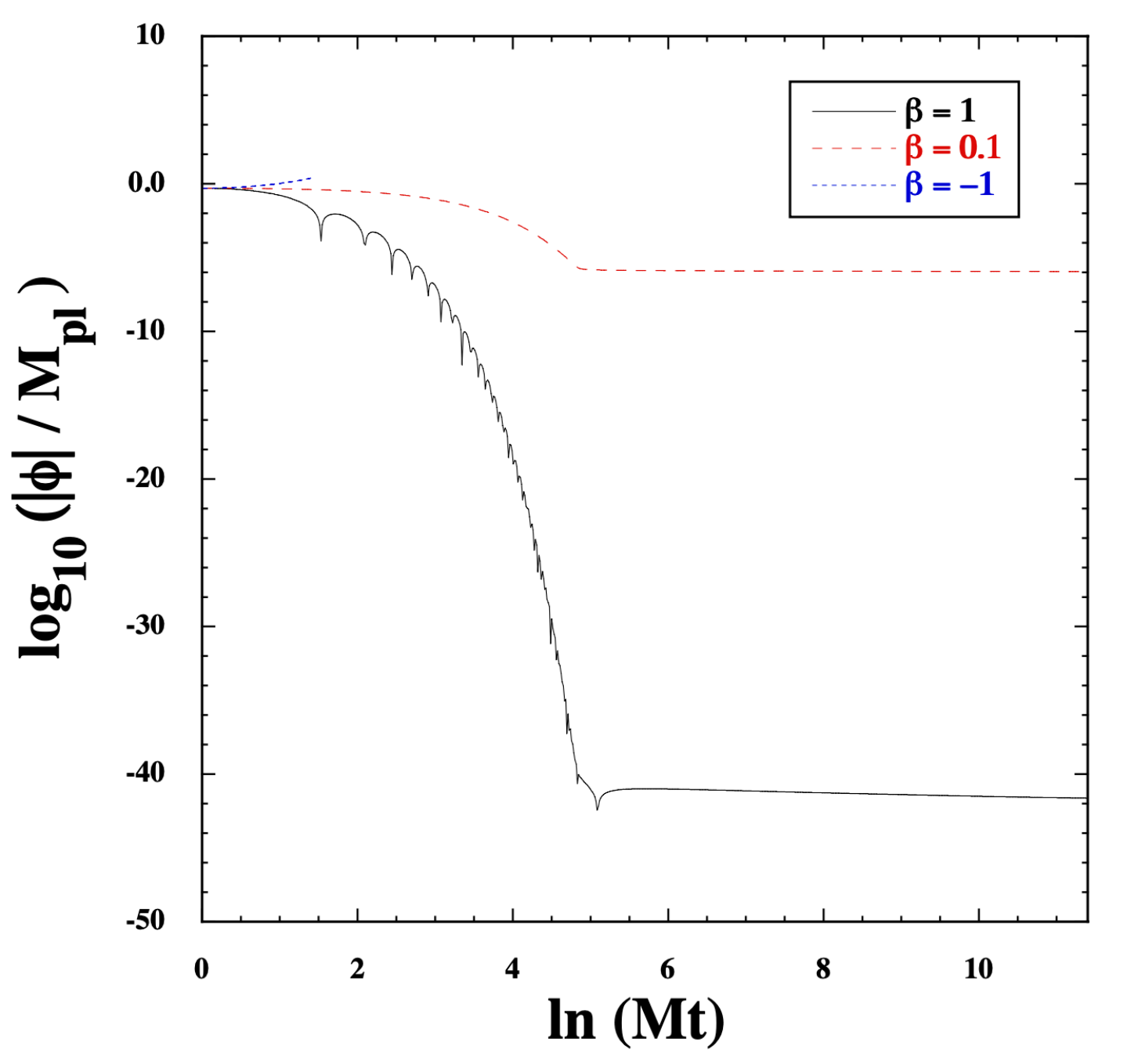}
\end{center}\vspace{-0.5cm}
\caption{\label{fig1}
Evolution of $|\phi|/\Mpl$ versus $\ln (Mt)$ during inflation 
and reheating for $\beta=1, 0.1, -1$ with $m=1.0 \times 10^{-11}$~eV 
and $M=1.1 \times 10^{-5}\Mpl$.
We choose the $\alpha$-attractor potential (\ref{Vchi}) 
with $\alpha=1$ and the decay constant $\Gamma=10^8$~GeV. 
The initial conditions of scalar fields are
$\chi_i=5.365 \Mpl$, 
$\dot{\chi}_i=0$, $\phi_i=0.5 \Mpl$, and $\dot{\phi}_i=0$. 
The integration is performed by the time at which the energy 
density of inflaton drops below that of radiation.
}
\end{figure}

In Fig.~\ref{fig1}, we plot the evolution of $|\phi|/\Mpl$ during inflation 
and reheating for $\beta=1, 0.1, -1$ with 
$m=10^{-11}$~eV and $M=10^{-5} \Mpl$. 
The initial conditions are chosen to be $\chi_i=5.365 \Mpl$, 
$\dot{\chi}_i=0$, $\phi_i=0.5 \Mpl$, and $\dot{\phi}_i=0$. 
For $\beta=1$, we can confirm that the amplitude of $\phi$ during inflation 
decreases as $|\phi| \propto \exp(-3Ht/2)$ with oscillations. 
In this simulation the number of e-foldings acquired during inflation 
is $N \simeq 60$, so the amplitude of $\phi$ at the end of inflation 
is of order $|\phi_f| \simeq |\phi_i| \exp(-90) \simeq 10^{-40}\Mpl$.
This rapid decrease of $\phi$ toward 0 is the outcome of a positive 
mass squared larger than $H^2$ induced by the nonminimal coupling 
with $\beta>3/8$.
Due to the strong suppression of $\phi$, the dynamics of inflation 
driven by the $\chi$-field potential energy $U(\chi)$ is 
not affected by the presence of $\phi$. 
For $\beta=0.1$, the analytic estimation (\ref{phidec}) shows that 
the field $\phi$ decreases as $|\phi| \propto \exp(-0.215Ht)$, 
so that $|\phi_f| \simeq 10^{-6}\Mpl$ at the end of inflation.
Even in this case, the dynamics of inflation is hardly modified by 
the field $\phi$. 

When $\beta=-1$, the field $\phi$ grows as $\phi \propto \exp(1.372Ht)$ 
from the onset of inflation, the slow-roll inflation is prevented 
by the rapid increase of $\dot{\phi}^2$ (see Fig.~\ref{fig1}).
In particular, the epoch of cosmic acceleration soon comes to end 
by the negative coupling $\beta \simeq -5$ used for 
the occurrence of spontaneous scalarization with $V(\phi)=0$. 
In our setup, the presence of the 
self-interacting potential $V(\phi)$ 
with a positive nonminimal coupling $\beta$ allows a possibility for 
realizing a positive effective field mass squared around $\phi=0$.
As discussed above, for $\beta>{\cal O}(0.1)$, the field $\phi$ decreases 
toward the local minimum of its effective potential ($\phi=0$) during inflation.

After inflation, the inflaton field $\phi$ should decay to radiation. 
To study the dynamics of $\phi$ during reheating, 
we incorporate the Born decay term $\Gamma \dot{\chi}$  
in Eq.~(\ref{chieq}) as 
\be
\ddot{\chi}+ \left( 3H +\Gamma \right) \dot{\chi}
+U_{,\chi}=0\,,
\label{dchieq}
\ee
where $\Gamma$ is a constant. 
The radiation density $\rho_r$ obeys the differential equation 
\be
\dot{\rho}_r
+4H \rho_r=\Gamma \dot{\chi}^2\,.
\label{rhor}
\ee
The energy density $\rho$ and pressure $P$ in Eqs.~(\ref{cosmo1}) and (\ref{cosmo2}) 
should be also modified to $\rho=\dot{\chi}^2/2+U(\chi)+\rho_r$ 
and $P=\dot{\chi}^2/2-U(\chi)+\rho_r/3$, respectively. 
We numerically solve Eqs.~(\ref{cosmo1})-(\ref{cosmo2}) and 
(\ref{dchieq})-(\ref{rhor}) by using the field values $\chi_f$, $\phi_f$, 
and their time derivatives at the end of inflation
as the initial conditions of the reheating period. 
We take the radiation into account from the end of inflation and integrate 
the background equations of motion by the time at which the inflaton energy 
density drops below $\rho_r$. 
For the mass $m$ of order $10^{-11}$~eV the condition $m^2 f_B^2 \ll H^2 \Mpl^2$ 
is satisfied in the standard 
reheating scenario, and it is a good approximation to neglect 
the contributions of the potential energy $V(\phi)$ to the background 
equations of motion.

The inflaton potential is approximated as $U(\chi) \simeq M^2 \chi^2/2$ 
around $\chi=0$. The reheating stage driven by the oscillating $\chi$ field 
corresponds to a temporal matter era with $a \propto t^{2/3}$ and $H=2/(3t)$. 
As long as the field $\phi$ sufficiently approaches $0$ during inflation, 
Eq.~(\ref{cosmo3}) approximately reduces to 
\be
\ddot{\phi}+\frac{2}{t} \dot{\phi}+\frac{2\beta}{3t^2}\phi \simeq 0\,.
\ee
The dominant solution to this equation is given by 
\begin{numcases}{\phi \propto }
     t^{-1/2} \cos \left(  \sqrt{\frac{8\beta-3}{12}}\ln (Mt)+\theta_0 \right)    &   
     ({\rm if}~$\beta>3/8)$\,, \label{damposre} \\
    t^{-(1-\sqrt{1-8\beta/3})/2} &   ({\rm if}~$\beta < 3/8)$\,.\label{phidecre}
\end{numcases}
The time $t_f$ at the beginning of reheating is related to the 
Hubble parameter at the end of inflation $H_f$, as $t_f \simeq 1/H_f$.
The reheating period ends around the time $t_R \simeq 1/\Gamma$, 
after which the energy density of radiation dominates over 
that of the inflaton field $\chi$. 
Since the evolution of $\phi$ during inflation is 
given by Eqs.~(\ref{dampos})-(\ref{phidec}), 
the amplitude of $\phi$ at which the radiation-dominated epoch 
commences can be estimated as 
\begin{numcases}{|\phi_R|=}
    |\phi_i| \exp \left(-\frac{3}{2}N \right) \left( \frac{\Gamma}{H_f} \right)^{1/2}&   
     ({\rm if}~$\beta>3/8)$\,, \label{phiR1} \\
     |\phi_i| \exp \left[ -\frac{3}{2} \left( 1-\sqrt{1-\frac{8}{3}\beta} 
\right)N \right] \left( \frac{\Gamma}{H_f} \right)^{( 1-\sqrt{1-8\beta/3})/2} 
 &   ({\rm if}~$0<\beta < 3/8)$\,,\label{phiR2}
\end{numcases}
where $\phi_i$ is the initial value of $\phi$ at the onset of inflation and 
$N$ is the total number of e-foldings during inflation. 
Since $\Gamma/H_f<1$, the amplitude of $\phi$ further decreases 
during the reheating epoch, but the suppression of $|\phi|$ is much 
less significant compared to the inflationary period. 
For $\beta>3/8$, $|\phi_R|$ does not depend on the 
coupling constant $\beta$.

The numerical simulation of Fig.~\ref{fig1} corresponds 
the decay constant $\Gamma=10^8$~GeV. 
The Hubble parameter around the end of inflation is of order 
$H_f=0.1M \simeq 10^{-6}M_{\rm pl}$. 
Applying the estimations (\ref{phiR1}) and  (\ref{phiR2}) 
to $\beta=1$ and $\beta=0.1$, we obtain 
$|\phi_{R}| \simeq 3 \times 10^{-42}\Mpl$ and 
$|\phi_{R}| \simeq 6 \times 10^{-7}\Mpl$, respectively,  
whose orders agree with the numerical results. 
For smaller $\Gamma$, the suppression of $|\phi|$ during 
reheating is even more significant. 
Thus, we showed that the positive nonminimal coupling 
with $\beta>{\cal O}(0.1)$ leads to the values of $\phi_R$ 
close to 0. This property is mostly attributed to the 
exponential decrease of $|\phi|$ during inflation.

\subsection{Evolution after the onset of radiation era}

Let us proceed to the discussion about the evolution of $\phi$ 
after the end of reheating by considering the 
mass scale of order $m={\cal O}(10^{-11}\,{\rm eV})$. 
During the radiation-dominated epoch, we have $H=1/(2t)$ and 
hence the term $3(2H^2+\dot{H})$ in Eq.~(\ref{cosmo3}) vanishes.
Provided that the field $\phi$ is much smaller than $\Mpl$ 
and $f_B$, Eq.~(\ref{cosmo3}) is 
approximately given by 
\be
\ddot{\phi}+3H \dot{\phi}-\left[ m^2 +
\frac{\beta (1+3\beta)\dot{\phi}^2}{2\Mpl^2} \right]\phi \simeq 0\,.
\label{sphieq}
\ee
For $\beta>3/8$ the initial field value (\ref{phiR1}) at the onset of the radiation 
era is as small as $10^{-42}\Mpl$, 
and we can ignore the second term in 
the square bracket of Eq.~(\ref{sphieq}) relative to $m^2$. 
For $0<\beta<3/8$, the field $\phi$ is not necessarily subject to 
strong suppression during inflation, so it may be possible 
to satisfy the condition $\beta  (1+3\beta)\dot{\phi}^2/(2\Mpl^2) \gg m^2$ 
at the end of reheating. 
In this case, however, $\dot{\phi}=0$ is the solution to 
Eq.~(\ref{sphieq}) and hence the field derivative rapidly decreases to 
reach the region $\beta  (1+3\beta)\dot{\phi}^2/(2\Mpl^2) \ll m^2$.
Thus, in both cases, the scalar field $\phi$ eventually obeys
\be
\ddot{\phi}+3H \dot{\phi}-m^2 \phi \simeq 0\,,
\label{sphiapeq}
\ee
which has a tachyonic mass squared $-m^2$ around $\phi=0$ 
due to the existence of the self-interacting
potential $V(\phi)$.
Since the condition $H \gg m={\cal O}(10^{-11}\,{\rm eV})$ is satisfied 
in the early radiation era, $\phi$ is nearly frozen by the Hubble friction. 

\subsubsection{Growth of the scalar field from 
the symmetry-restored state}

After $H$ drops below the order $m$, $\phi$ starts to increase. 
During the radiation dominance, the solution to Eq.~(\ref{sphiapeq}) 
is given by 
\be
\phi=t^{-1/4} \left[ c_1I_{1/4} (mt)+c_2 K_{1/4} (mt) 
\right]\,,
\label{phirad}
\ee
where $I_{1/4}$ and $K_{1/4}$ are modified Bessel functions of 
the first and second kinds, respectively. 
Taking the limit $m t \gg 1$ in Eq.~(\ref{phirad}), there is indeed 
a growing-mode solution $\phi \propto {\rm e}^{mt}/t^{3/4}$. 
Since the potential (\ref{Vphi}) has a local minimum at $\phi=\pi f_B$, 
the field $\phi$ eventually reaches this region and starts to oscillate 
around $\phi=\pi f_B$.

\begin{figure}[ht]
\begin{center}
\includegraphics[height=3.2in,width=3.4in]{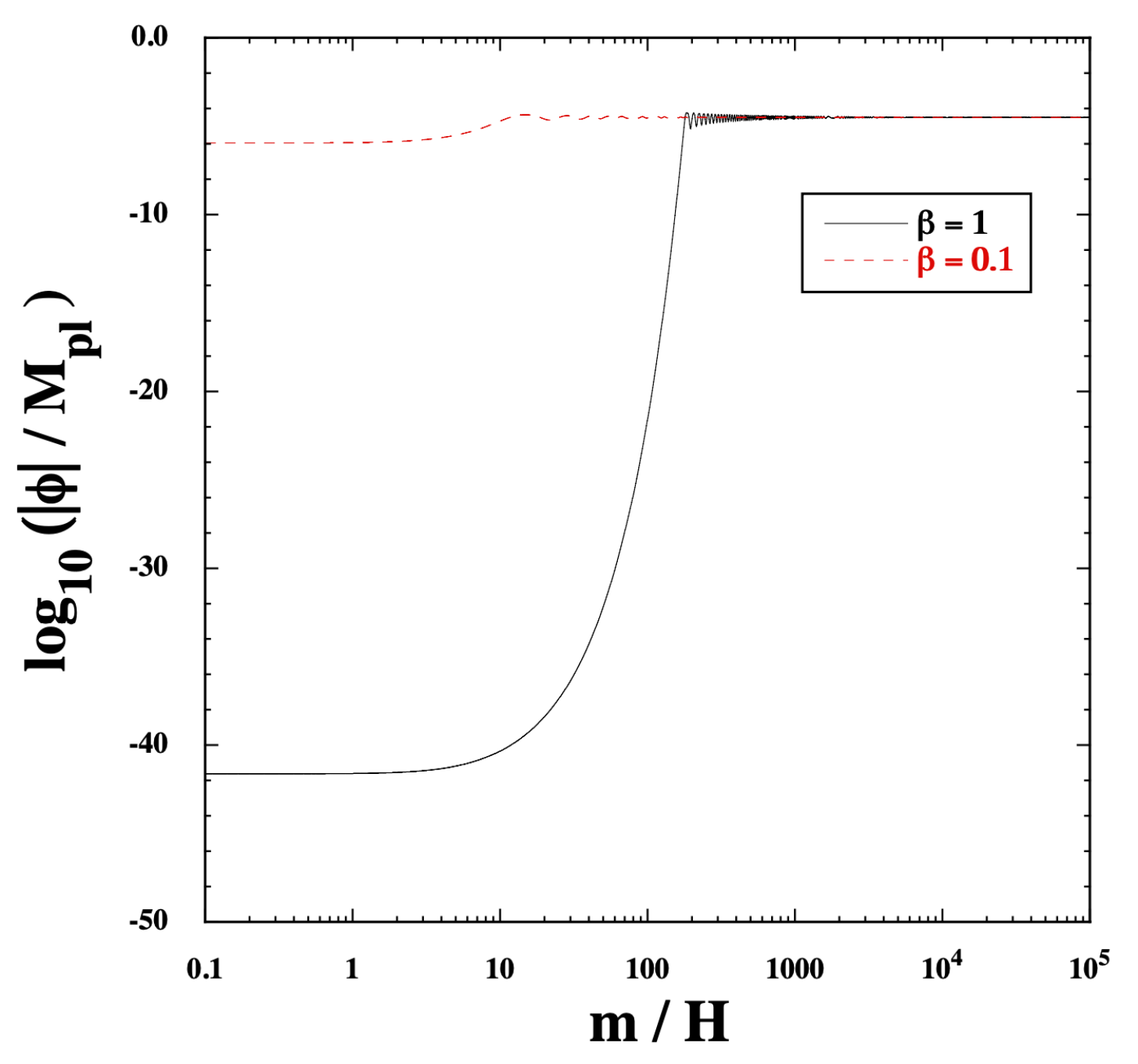}
\end{center}\vspace{-0.5cm}
\caption{\label{fig2}
Evolution of $|\phi|/\Mpl$ versus $m/H$ in the radiation-dominated 
epoch for $\beta=1, 0.1$ with $m=1.0 \times10^{-11}$~eV and $f_B=1.0 \times 10^{-5}\Mpl$. 
The initial conditions of $\phi$ are chosen to match with those derived 
in the numerical simulations of Fig.~\ref{fig1} at the end of reheating.
}
\end{figure}

In Fig.~\ref{fig2}, we plot the evolution of $|\phi|/\Mpl$ as a function of $m/H$ 
for $\beta=1, 0.1$. We choose $m=10^{-11}$~eV and $f_B=1.0 \times 10^{-5}\Mpl$ 
with the initial conditions of $\phi$ consistent with their values at the end 
of reheating. For $\beta =1$, the field $\phi$ is nearly frozen with 
the value of order $10^{-42}\Mpl$ and then starts to grow for $H\lesssim m/3$.
Around $H\lesssim m/200$, the field sufficiently approaches the 
potential minimum and exhibits a damped oscillation around $\phi=\pi f_B$.
When $\beta=0.1$, the field starts to evolve for $H\lesssim m/3$ as well, 
while the approach to $\phi=\pi f_B$ occurs around $H\lesssim m/12$
because of the large initial value of $\phi$ of order $10^{-6}\Mpl$.

In the era dominated by the radiation density $\rho_r=\pi^2 g_* T^4/30$, 
where $g_*$ is the number of relativistic degrees of freedom 
and $T$ is the temperature,  
we estimate the temperature $T_m$ at which the field $\phi$ 
starts to evolve along the potential $V(\phi)$. 
Using the Friedmann equation $3H^2 \Mpl^2=\rho_r$
with $m=3H$, it follows that 
\be
T_m=\left( \frac{10}{\pi^2 g_*} \right)^{1/4} 
\sqrt{m \Mpl}\,.
\ee
For the mass scale $m=10^{-11}$~eV with $g_* \simeq 10$ \cite{Dodelson:2003ft}, 
we have $T_m \simeq 10^{12}$~K. 
Then, the field $\phi$ reaches 
the potential minimum $\pi f_B$ before the epoch of
big-bang nucleosynthesis (BBN). 
After the Universe enters the matter-dominated epoch, 
the term $\dot{H}+2H^2$ in Eq.~(\ref{cosmo3}) is nonvanishing, 
i.e., $\dot{H}+2H^2 \simeq H^2/2$. 
Since $m^2 \gg H^2$ in this epoch, the effect of the nonminimal 
coupling on Eq.~(\ref{cosmo3}) is negligible and the field 
$\phi$ coherently oscillates around $\phi=\pi f_B$ 
with a decreasing amplitude.
This is also the case for the late-time dark energy 
dominated era, so the field $\phi$ reaches
the potential minimum by today.

\subsubsection{Oscillation around the potential minimum as CDM}

Around $\phi=\pi f_B$, the potential (\ref{Vphi}) is approximated 
as 
\be
\label{osc_potential}
V(\phi) \simeq
\frac{1}{2}
m^2 (\phi-\pi f_B)^2\,.
\ee
After the scalar field starts to oscillate about the potential minimum,  
it behaves as (a portion of)
CDM with the energy density 
decreasing as $\rho_\phi \propto a^{-3}$. 
Today's field density can be estimated as 
$\rho_{\phi0} \simeq m^2 f_B^2 a_{\rm CDM}^3$, where 
$a_{\rm CDM}$ is the scale factor at which the field $\phi$ 
starts to behave as CDM during the radiation era 
and the scale factor is normalized as $a_0=1$.
Defining the ratio 
\be
r_{\rm CDM}=\frac{m}{H_{\rm CDM}}
\ee
with $H_{\rm CDM}=H_0 \sqrt{\Omega_{r0}a_{\rm CDM}^{-4}}$, 
where $H_0$ and $\Omega_{r0}$ are today's Hubble parameter and 
radiation density parameter respectively, today's field density parameter 
$\Omega_{\phi 0}=\rho_{\phi 0}/(3\Mpl^2 H_0^2)$ can be estimated as
\be
\Omega_{\phi 0}=\frac{r_{\rm CDM}^{3/2}}{3} 
\left( \frac{f_B}{\Mpl} \right)^2 \left( \frac{m}{H_0} 
\right)^{1/2} \Omega_{r0}^{3/4}\,.
\ee
If the field $\phi$ is responsible for a part of CDM, 
we require that 
$\Omega_{\phi 0} \simeq 0.27 \alpha_{\rm CDM}$,
where the constant $\alpha_{\rm CDM}$ represents the energy 
fraction of $\phi$ to CDM and $\alpha_{\rm CDM}=1$ 
corresponds to the case that $\phi$ is responsible 
for all CDM. 
Then, we obtain
\be
\frac{f_B}{\Mpl} \simeq 30 r_{\rm CDM}^{-3/4}
\sqrt{\alpha_{\rm CDM}}
\left( \frac{m}{10^{-33}~{\rm eV}} 
\right)^{-1/4} \,,
\label{fBcon}
\ee
where we used $\Omega_{r0} \simeq 9 \times 10^{-5}$ 
and $H_0 \simeq 10^{-33}$~eV.
If $m=10^{-11}$~eV, then we have 
$f_B/\Mpl \simeq 9.4 \times 10^{-5} 
\sqrt{\alpha_{\rm CDM}}\,r_{\rm CDM}^{-3/4}$. 
Using the value $r_{\rm CDM}=200$ for $\beta=1$, 
we obtain $f_B/\Mpl \simeq 2 \times 10^{-6} \sqrt{\alpha_{\rm CDM}}$.
For $\beta=0.1$ we take the value $r_{\rm CDM}=12$, 
in which case $f_B/\Mpl \simeq 1 \times 10^{-5}
\sqrt{\alpha_{\rm CDM}}$.
Under the constraint (\ref{fBcon}), the density parameter of 
$\phi$ at $a=a_{\rm CDM}$ (which is slightly before the BBN epoch) 
is as small as 
\be
\Omega_\phi (a=a_{\rm CDM})
\simeq 
\frac{m^2f_B^2}{3\Mpl^2 H_{\rm CDM}^2}
\simeq 
\frac{f_B^2}{3\Mpl^2}
r_{\rm CDM}^2
\simeq 
300
r_{\rm CDM}^{1/2}
\alpha_{\rm CDM}
\left(
\frac{m}
{10^{-33}{\rm eV}}
\right)^{-1/2}
=
3\times 10^{-9}
r_{\rm CDM}^{1/2} \alpha_{\rm CDM}
\,,
\ee
and hence the BBN is not affected by the presence of the field $\phi$.

The relation (\ref{fBcon}) has been derived by assuming that the scalar 
field $\phi$ behaves as a coherently oscillating CDM by today. 
If the energy density of $\phi$ decays to that of radiation or 
some other particle whose density decreases faster than radiation, 
then it is possible to have larger values of $f_B$ than those constrained by Eq.~(\ref{fBcon}). For example, adding a decay term 
$\Gamma_{\phi} \dot{\phi}$ to the left hand-side of Eq.~(\ref{cosmo3}) 
leads to the dissipation of the energy density of $\phi$ 
before the field reaches the VEV $\phi_v=\pi f_B$.
When we study scalarized NS solutions in Sec.~\ref{scasec}, 
we will allow for the possibility that $f_B/\Mpl$ is 
larger than the value constrained by Eq.~(\ref{fBcon}). 

\section{Weak gravitational objects}
\label{weaksec}

In this section, we study solutions of the scalar field $\phi$ for compact 
objects on weak gravitational backgrounds~(like the Sun).
For this purpose, we consider a static and spherically symmetric 
background given by the line element
\be
\rd s^2
=
g_{\mu\nu}
{\rm d}x^\mu {\rm d}x^\nu
=-f(r) \rd t^{2} +h^{-1}(r) \rd r^{2}
+ r^{2} \left(\rd \theta^{2}
+\sin^{2}\theta\,\rd\varphi^{2} 
\right)\,,
\label{BGmetric}
\ee
where $f(r)$ and $h(r)$ are functions of the radial
coordinate $r$. 
The scalar field is assumed to be a function of $r$ alone, 
i.e., $\phi=\phi(r)$. 
For the matter species inside a star, we consider a perfect 
fluid described by the mixed energy-momentum tensor 
$T^{\mu}{}_{\nu}={\rm diag} \left[ -\rho(r), P(r), P(r), P(r) \right]$, 
where the energy density $\rho$ and pressure $P$ are
functions of $r$. 
Assuming that the perfect fluid is minimally coupled to gravity, 
it obeys the continuity Eq.~\eqref{eq_continuity}.
On the background (\ref{BGmetric}), this equation translates to 
\be
P'+\frac{f'}{2f} \left( \rho+P \right)=0\,,
\label{mattereq}
\ee
where a `prime' represents the derivative with respect to $r$.

Varying the action (\ref{action}) with respect to $f$ and $h$, 
we obtain the following gravitational field equations
\ba
& &
\frac{f'}{f}=\frac{4(1-h)\Mpl^4+2\Mpl^2 hr \phi' 
(r \phi'+4\beta \phi)-3\beta^2 \phi^2 \phi'^2 h r^2
-4(V-P)r^2 \Mpl^2 {\rm e}^{\beta \phi^2/(2\Mpl^2)}}
{2\Mpl^2 hr (2\Mpl^2-\beta \phi \phi' r)}\,,
\label{feq}\\
& & 
\frac{h'}{h}-\frac{f'}{f}=\frac{r[2\Mpl^2 \beta h\phi \phi''
+\{ 2(\beta-1)\Mpl^2+\beta^2 \phi^2 \}h \phi'^2
-2\Mpl^2 (\rho+P){\rm e}^{\beta \phi^2/(2\Mpl^2)}]}
{\Mpl^2 h (2\Mpl^2-\beta \phi \phi' r)}\,.
\label{heq}
\ea
The scalar field obeys the differential equation 
\ba
& &
\phi''+\frac{2(1+h)\Mpl^2+\beta \phi'^2 h r^2
-r^2(2V+\rho-P){\rm e}^{\beta \phi^2/(2\Mpl^2)}}{2\Mpl^2 hr}\phi'
\nonumber \\
& &
-\frac{(2\Mpl^2-\beta \phi \phi' r){\rm e}^{\beta \phi^2/(2\Mpl^2)}}{4 \Mpl^4 h}
\left[ 2V_{,\phi}\Mpl^2+4\beta \phi V+\beta 
(\rho-3P) \phi \right]=0\,.
\label{phieqs}
\ea
As we discussed in Sec.~\ref{cossec}, for 
$\beta>{\cal O}(0.1)$, the scalar field $\phi$ approaches 
the VEV $\phi_v=\pi f_B$ by today during the cosmic expansion history.
We would like to construct a scalar-field profile $\phi(r)$ 
having the asymptotic value $\pi f_B$ at spatial infinity, i.e., 
\be
\phi (\infty)=\pi f_B\,,
\ee
with $\phi'(\infty)=0$.
At $r=0$, we impose the boundary conditions 
$\phi(0)=\phi_0$ and $\phi'(0)=0$, 
where $\phi_0$ is a constant in the range $0<\phi_0<\pi f_B$. 

Since we are now considering a nonrelativistic object, we ignore 
$P$ relative to $\rho$ and employ the approximation 
$\rho r^2 \ll \Mpl^2$ inside the star. 
The gravitational potentials are much smaller than 1, so we can 
exploit the approximation $h \simeq 1$ in Eq.~(\ref{phieqs}).
As we will see below, the field variation $\phi'(r)$ is small 
on weak gravitational backgrounds, under which 
$\beta \phi'^2 r^2 \ll \Mpl^2$ and 
$| \beta \phi \phi' r| \ll \Mpl^2$. 
In the vicinity of $\phi=\pi f_B$, the potential (\ref{Vphi}) can be 
also expanded as Eq.~\eqref{osc_potential}.
Around $\phi=\pi f_B$, Eq.~(\ref{phieqs}) is 
approximately given by 
\be
\phi''+\frac{2}{r}\phi'-\left[ m^2 \left( \phi-\pi f_B 
\right)+\frac{\beta \rho}{2\Mpl^2}\phi \right] \simeq 0\,.
\label{phiweak}
\ee
By the end of this section, we consider a star with the constant density 
$\rho$ and radius $r_s$. We assume that the exterior region of 
the star has a vanishing density ($\rho=0$). 
Then, for $r>r_s$, the solution to Eq.~(\ref{phiweak}) consistent with 
the boundary condition $\phi'(\infty)=0$ is given by 
\be
\phi(r)=\pi f_B+\frac{A{\rm e}^{-m r}}{r}\,,
\label{phiso1}
\ee
where $A$ is a constant.

Inside the star, the field Eq.~(\ref{phiweak}) can be expressed as
\be
\phi''+\frac{2}{r}\phi'-\mu^2 \left( \phi-\phi_0 
\right)=0\,,
\label{phiweak2}
\ee
where 
\be
\mu^2 \equiv m^2+\frac{\beta \rho}{2\Mpl^2}\,,\qquad 
\phi_0 \equiv \frac{m^2}{\mu^2}\pi f_B\,.
\label{muphi0}
\ee
For $\beta>0$, we have $\mu^2>m^2$ and hence $\phi_0<\pi f_B$.
If we consider the Sun with the mean density
$\rho={\cal O}(1~{\rm g/cm}^3)$ 
and mass $m={\cal O}(10^{-11}\,{\rm eV})$, 
we have $m^2 \gg \beta \rho/(2\Mpl^2)$, 
under which $\phi_0$ is very close to $\pi f_B$. 
Since we are assuming that $\rho={\rm constant}$,
the solution to Eq.~(\ref{phiweak2}) consistent with 
the boundary condition $\phi'(0)=0$ is given by 
\be
\phi(r)=\phi_0+\frac{B ({\rm e}^{\mu r}
-{\rm e}^{-\mu r})}{r}\,,
\label{phiso2}
\ee
where $B$ is a constant. 

Matching Eq.~(\ref{phiso1}) with (\ref{phiso2}) and also their 
$r$ derivatives at $r=r_s$, we obtain 
\be
A=
\frac{(\phi_0-\pi f_B)[(\mu r_s-1)\,{\rm e}^{2\mu r_s}+\mu r_s+1]
\,{\rm e}^{m r_s}}{(\mu+m)\,{\rm e}^{2\mu r_s}+\mu-m}\,,
\qquad 
B=
-\frac{(\phi_0-\pi f_B)(m r_s+1)\,{\rm e}^{m r_s}}{(\mu+m)
\,{\rm e}^{2\mu r_s}+\mu-m}\,.
\ee
Then, the resulting solution of $\phi$ outside the star ($r>r_s$) 
 is given by 
\be
\phi(r)=\pi f_B -\beta_{\rm eff} \Mpl \frac{GM_g}{r} 
{\rm e}^{-m (r-r_s)} \,,
\label{phica}
\ee
where $G=1/(8 \pi \Mpl^2)$ is the gravitational constant, 
$M_g=4\pi r_s^3 \rho/3$ is the mass of body, and 
\be
\beta_{\rm eff}=3 \beta \frac{\pi f_B}{\Mpl}
\frac{(\mu r_s-1)\,{\rm e}^{2\mu r_s}+\mu r_s+1}
{\mu^2 r_s^3 [(\mu+m)\,{\rm e}^{2\mu r_s}+\mu-m]}\,.
\label{beff}
\ee
The fifth force between the scalar field and baryons 
is mediated by the effective coupling $\beta_{\rm eff}$. 
Note that the solution (\ref{phica}) looks similar to that
derived in the chameleon mechanism \cite{Khoury:2003aq,Khoury:2003rn}, 
but the difference is that the effective mass of $\phi$ inside and 
outside the star 
is similar to each other ($\mu \simeq m$) in our scenario. 
This leads to the different form of $\beta_{\rm eff}$ 
in comparison to the chameleon case.

If we consider the Sun ($r_s=7.0 \times 10^8$~m) with the mass 
$m=10^{-11}$~eV, we have $\mu^2 \simeq m^2 \gg \beta \rho/(2\Mpl^2)$ 
and $\mu r_s \simeq m r_s \simeq 3.5 \times 10^4$.
In this case, Eq.~(\ref{beff}) reduces to 
\be
\beta_{\rm eff} \simeq \frac{3\beta}{2} 
\frac{\pi f_B}{\Mpl}\frac{1}{(mr_s)^2} 
\qquad {\rm for} \quad 
mr_s \gg 1\,.
\label{beeff}
\ee
Because of a large suppression factor $(mr_s)^{-2}$, it is
easier to satisfy Solar-system constraints in comparison 
to the massless case (see below).
For the symmetry-breaking 
scale $f_B/\Mpl=2 \times 10^{-6}$ 
with $\beta=1$, the effective coupling is as small as 
$\beta_{\rm eff} \simeq 7.7 \times 10^{-15}$.
In the case of Earth ($r_s=6.4 \times 10^6$~m) 
with $m=10^{-11}$~eV, $f_B/\Mpl=2 \times 10^{-6}$, 
and $\beta=1$, we have $\beta_{\rm eff} \simeq 9.2 \times 10^{-11}$.
In addition to these small values of $\beta_{\rm eff}$, 
the factor ${\rm e}^{-m (r-r_s)}$ in Eq.~(\ref{phica}) leads to the exponential
suppression of fifth forces at the distance $r \gtrsim r_s+1/m$. 

In the massless limit $m r_s \to 0$, we have $\mu^2 \simeq \beta \rho/(2\Mpl^2)$. 
Since $(\mu r_s)^2$ is of order the gravitational potential at the surface of star, 
we exploit the approximation $\mu r_s \ll 1$ in Eq.~(\ref{beff}). 
Then, the effective coupling reduces to 
\be
\beta_{\rm eff} \simeq \beta \frac{\pi f_B}{\Mpl}
\qquad {\rm for} \quad 
mr_s \to 0\,.
\label{beeff2}
\ee
For $\beta$ of order 1, we have $\beta_{\rm eff} \ll 1$ 
under the condition $\pi f_B/\Mpl \ll 1$, 
in which case it is possible to satisfy Solar-system constraints 
even for a nearly massless scalar field (as we will see 
at the end of this section).

Outside the star, we estimate fifth-force corrections to the metric components 
$f$ and $h$. 
They are related to
the gravitational potentials $\Psi$ and $\Phi$, 
as $f={\rm e}^{2\Psi}$ and $h={\rm e}^{2\Phi}$. 
Since $|\Psi|$ and $|\Phi|$ are much smaller than 1 on 
weak gravitational backgrounds (of order $10^{-6}$ for the Sun), 
we only pick up terms linear in $\Psi$ and $\Phi$. 
Let us consider the massive scalar field satisfying the condition 
\be
m r_s \gg 1\,.
\ee
Substituting the solution (\ref{phica}) and its $r$ derivatives 
into Eqs.~(\ref{feq}) and (\ref{heq}), we find that the gravitational potentials 
$\Phi$ and $\Psi$ approximately obey
\ba
& &
\Phi'+\frac{\Phi}{r} \simeq -\frac{1}{2} \beta \beta_{\rm eff} G M_g 
\frac{\pi f_B}{\Mpl} m^2 {\rm e}^{-m (r-r_s)}\,,\\
& &
\Psi'+\frac{\Phi}{r} \simeq \beta \beta_{\rm eff} \frac{G M_g}{r} 
\frac{\pi f_B}{\Mpl} m\,{\rm e}^{-m (r-r_s)}\,.
\ea
The integrated solutions to these equations, which are consistent with 
the asymptotic flatness, are given by 
\ba
\Phi &=& -\frac{G M_g}{r} \left[ 1-\frac{\beta \beta_{\rm eff}}{2}
\frac{\pi f_B}{\Mpl}mr\,{\rm e}^{-m (r-r_s)} \right]\,,\label{Phi}\\
\Psi &=&  -\frac{G M_g}{r} \left[ 1+\frac{\beta \beta_{\rm eff}}{2}
\frac{\pi f_B}{\Mpl} {\rm e}^{-m (r-r_s)} \right]\,.\label{Psi}
\ea
We introduce the post-Newtonian parameter as 
\be
\gamma_{\rm PPN} \equiv \frac{\Phi}{\Psi}
 \simeq 1-\frac{\beta \beta_{\rm eff}}{2}
\frac{\pi f_B}{\Mpl}mr\,{\rm e}^{-m (r-r_s)}\,.
\ee
The time-delay effect of the Cassini tracking of the Sun 
has given the bound $\gamma_{\rm PPN}-1=
(2.1 \pm 2.3) \times 10^{-5}$ \cite{Will:2014kxa}. 
Since $\gamma_{\rm PPN}-1$ is negative in the current theory, 
we adopt the limit $1-\gamma_{\rm PPN} \leq 2.0 \times 10^{-6}$.
Taking the value of $\gamma_{\rm PPN}$ at $r=r_s$, 
this Solar-system constraint translates to
\be
\beta \beta_{\rm eff}\frac{\pi f_B}{\Mpl}mr_s \le 
4.0 \times 10^{-6}\,.
\label{bebound0}
\ee
On using the effective coupling (\ref{beeff}) derived for 
$m r_s \gg 1$, we obtain the bound 
\be
\beta \frac{\pi f_B}{\Mpl} \frac{1}{\sqrt{mr_s}} \le 
1.6 \times 10^{-3}\qquad {\rm for} \quad 
mr_s \gg 1\,.
\label{bebound}
\ee
With the mass scale $m=10^{-11}$~eV, this translates to 
$\pi f_B/\Mpl \leq 0.3/\beta$ for the Sun.
The symmetry-breaking scale $f_B/\Mpl \simeq 2 \times 10^{-6}$ with $\beta=1$, which was mentioned in Sec.~\ref{cossec} 
in the context of an oscillating $\phi$-field CDM, is well 
consistent with this upper limit.

We also comment on Solar-system 
constraints in the massless limit ($m r_s \to 0$). 
In this case, the scalar-field solution is given by Eq.~(\ref{phica}) 
with $\beta_{\rm eff}=\beta \pi f_B/\Mpl$. 
Provided that $\pi f_B/\Mpl$ is smaller than the order 1, 
the gravitational potential $\Phi$ is estimated as 
$\Phi=-G M_g/r$ up to the linear order in $GM_g/r$, 
while the other gravitational potential receives 
a correction from the nonminimal coupling 
as $\Psi=-(GM_g/r)[1+\beta^2 (\pi f_B/\Mpl)^2]$. 
Then, the post-Newtonian parameter is estimated as 
\be
\gamma_{\rm PPN} \simeq 1-\beta^2 
\left( \frac{\pi f_B}{\Mpl} \right)^2\,,
\label{gammassless}
\ee
where we used the approximation $\beta^2 (\pi f_B/\Mpl)^2 \ll 1$. 
Note that the result (\ref{gammassless}) is consistent with that  
derived in Ref.~\cite{Damour:1992we}. 
Using the Solar-system bound
$1-\gamma_{\rm PPN} \leq 2.0 \times 10^{-6}$, 
it follows that 
\be
\beta \frac{\pi f_B}{\Mpl} \le 1.4 \times 10^{-3} 
\qquad {\rm for}\quad 
mr_s \to 0\,.
\label{bebound2}
\ee
For the symmetry-breaking 
scale $f_B/\Mpl \simeq 2 \times 10^{-6}$ 
with $\beta=1$, the bound (\ref{bebound2}) is satisfied. 
In the massive case (\ref{bebound}) there is an extra 
suppression factor $1/\sqrt{m r_s} \ll 1$, 
and the propagation 
of fifth forces is suppressed in comparison to 
the massless case.  

For laboratory tests of gravity, the associated scale $r$ of experiments 
is in the range $m r \ll 1$. 
Let us consider two identical test bodies with constant density 
$\rho$, radius $r_s$, and mass $M_g=4\pi r_s^3 \rho/3$. 
In this case, the gravitational potential $\Psi$ made by one test body 
is given by Eq.~(\ref{Psi}), with $\beta_{\rm eff} \simeq 
\beta \pi f_B/\Mpl$ and ${\rm e}^{-m(r-r_s)} \simeq 1$. 
Then, the potential energy between two test bodies is expressed as
\be
V(r)=M_g \Psi=-\frac{G M_g^2}{r} \left[1+\frac{\beta^2}{2}
\left( \frac{\pi f_B}{\Mpl} \right)^2 \right]\,.
\label{Vr}
\ee
The second term in the squared bracket of Eq.~(\ref{Vr}), 
which corresponds to the fifth-force contribution to $V(r)$, 
can be expressed in the form $\beta_{\rm eff}^2/2$. 
This result is analogous to what was obtained for chameleon 
and symmetron theories \cite{Khoury:2003rn,Sakstein:2017pqi}.
Parametrizing the fifth-force potential energy as 
$V_{\rm f}(r)=-\alpha_{\rm f} GM_g^2/r$, the laboratory tests of gravity 
gives the constraint $\alpha_{\rm f}<10^{-3}$ \cite{Adelberger:2009zz}.
Since $\alpha_{\rm f}=(\beta^2/2)(\pi f_B/\Mpl)^2$ in our case, 
we obtain the following bound
\be
\beta \frac{\pi f_B}{\Mpl} \le 4.5 \times 10^{-2}\,.
\label{bebound3}
\ee
This is weaker than the Solar-system constraint (\ref{bebound2}) 
by one order of magnitude.
On astrophysical scales much larger than $m^{-1}={\cal O} (10\,{\rm km})$, 
our model is consistent with observational tests of gravity  
due to the exponential suppression of fifth forces.

For the choice $m^{-1}={\cal O} (10\,{\rm km})$, we may also apply the experimental tests of Newton's law on geophysical scales to our model.
Assuming that the Yukawa-type corrections to the Newtonian potential
$V_g(r)=-(GM_g/r) \left(1+\alpha_{\rm f} e^{-r/\lambda}\right)$ 
for the scale $\lambda={\cal O} (10\,{\rm km})$, the bound on 
$\alpha_{\rm f}$ is given by $|\alpha_{\rm f}| \lesssim 10^{-4}$ 
(see e.g., Fig.~4 of Ref.~\cite{Adelberger:2003zx}).
Comparing $V_g(r)$ with Eq. \eqref{Psi}, we find 
\be
\beta 
\beta_{\rm eff}
\frac{\pi f_B}{\Mpl}
\lesssim 10^{-4}\,.
\label{begeo}
\ee
This is weaker than the Solar System bound \eqref{bebound0}
derived for $m r_s \gg 1$. 

\section{Neutron star solutions}
\label{scasec}

In this section, we will construct NS solutions on the static and 
spherically symmetric background given by the line element 
(\ref{BGmetric}). We note that Eqs.~(\ref{mattereq})-(\ref{phieqs}) 
are the strict Euler-Lagrange equations obtained by varying the action \eqref{action} and hence valid also on strong gravitational backgrounds.
The difference from the case of weak gravitational stars
discussed in Sec.~\ref{weaksec} is that the gravitational 
potentials $|\Psi|$ and $|\Phi|$ in the vicinity of NSs 
are of ${\cal O}(0.1)$ and nonlinearities in the gravitational 
field equations become important.
Moreover, the pressure $P$ is not negligible 
relative to the energy density $\rho$.
The other important difference is that the central density 
of NSs $\rho_c$ is typically of ${\cal O}(10^{15}~{\rm g/cm}^3)$,
so in our model the term $\beta \rho/(2\Mpl^2)$ can exceed 
$m^2={\cal O}\left( (10^{-11}~{\rm eV})^2\right)$ 
for $\beta \gtrsim {\cal O}(0.1)$. 
This means that the field value $\phi_0$ defined in Eq.~(\ref{muphi0}) can 
approach 0 inside the NS, unlike the low density star 
where $\phi_0$ is very close to $\pi f_B$. 
Then, it should be possible to realize a scalar-field configuration 
in which $\phi$ is close to $\phi=0$ inside the star and 
approaches $\pi f_B$ outside the star. 
In the following, we will show that such scalarized solutions do exist.

\subsection{Boundary conditions}

We first derive the approximate solutions around the center of star 
by using the expansions of $f$, $h$, $\phi$, and $P$.
Due to the regularity condition $\phi'(0)=0$, we can expand 
the scalar field around $r=0$ in the form 
$\phi (r)=\phi_c+\phi_2 r^2+{\cal O}(r^3)$, where 
$\phi_c=\phi(0)$ and
$\phi_2$ is a constant.
We also impose the boundary conditions 
$f(0)=f_c$, $h(0)=1$, $\rho(0)=\rho_c$, $P(0)=P_c$, and 
$f'(0)=h'(0)=\rho'(0)=P'(0)=0$. Around $r=0$, 
the scalar-field potential is expanded as
\be
V(\phi)=V_c+V_{,\phi c} (\phi-\phi_c)
+{\cal O} ((\phi-\phi_c)^2)\,,
\ee
where we used the notations $V_c \equiv V(\phi_c)$ 
and $V_{,\phi c} \equiv (\rd V/\rd \phi)(\phi_c)$.
The solutions consistent with Eqs.~(\ref{mattereq})-(\ref{phieqs}) 
around the center of NSs are given by 
\ba
f &=& f_c+f_c \frac{{\rm e}^{\beta \phi_c^2/(2\Mpl^2)}
[2(\rho_c+3P_c-2V_c)\Mpl^2+2\beta \phi_c \Mpl^2 V_{,\phi c}
+\beta^2 \phi_c^2 (\rho_c-3P_c+4V_c)]}{12\Mpl^4}r^2
+{\cal O}(r^4)\,,\label{fr=0} \\
h &=& 1-\frac{{\rm e}^{\beta \phi_c^2/(2\Mpl^2)}
[2(\rho_c+V_c)\Mpl^2-2\beta \phi_c \Mpl^2 V_{,\phi c}
-\beta^2 \phi_c^2 (\rho_c-3P_c+4V_c)]}{6\Mpl^4}r^2
+{\cal O}(r^4)\,,\\
\phi &=&\phi_c+\frac{{\rm e}^{\beta \phi_c^2/(2\Mpl^2)}}{6} 
\left[ V_{,\phi c}+\frac{\beta \phi_c (\rho_c-3P_c+4V_c)}{2\Mpl^2} 
\right]r^2+{\cal O}(r^4)\,,\label{phir=0}\\
P &=& P_c-\frac{{\rm e}^{\beta \phi_c^2/(2\Mpl^2)} (\rho_c+P_c)
[2(\rho_c+3P_c-2V_c) \Mpl^2+2\beta \phi_c \Mpl^2 V_{,\phi c}
+\beta^2 \phi_c^2 (\rho_c-3P_c+4V_c)]}{24\Mpl^4}r^2
+{\cal O}(r^4)\,.
\label{Pr=0} 
\ea
Let us consider the case in which $\phi_c$ is in the range 
$0<\phi_c \ll \pi f_B$. 
The potential energy around $\phi=0$ is of order 
$V \simeq 2m^2 f_B^2$, with $V_{,\phi} \simeq -m^2 \phi$. 
Provided that $f_B \ll \Mpl$, it follows that $V$ is much smaller than 
the central density $\rho_c={\cal O}(10^{15}\,{\rm g/cm}^3)$ 
for $m ={\cal O}( 10^{-11}$~eV).
Then, the solution (\ref{phir=0}) is approximately 
given by 
\be
\phi \simeq \phi_c+\frac{{\rm e}^{\beta \phi_c^2/(2\Mpl^2)}}
{6}\phi_c m_{\rm eff}^2 r^2+{\cal O}(r^3)\,,
\label{phiap}
\ee
where 
\be
m_{\rm eff}^2 \equiv -m^2+\frac{\beta \rho_c(1-3w_c)}{2\Mpl^2}\,,
\label{meffs}
\ee
with the equation of state (EOS) parameter $w_c=P_c/\rho_c$ 
at $r=0$. Here, $m_{\rm eff}^2$ corresponds to an effective mass squared 
of the scalar field around the potential maximum at $\phi=0$.
Like Eq.~(\ref{m0}), for $\beta=0$, we have $m_{\rm eff}^2=-m^2<0$, 
so the scalar field decreases as a function of $r$, i.e., $\phi'(r)<0$, 
around $r=0$. In the presence of the positive nonminimal coupling $\beta$ 
with $w_c<1/3$, it is possible to realize $m_{\rm eff}^2>0$ for 
\be
\beta>\frac{2m^2 \Mpl^2}{\rho_c (1-3w_c)}
=\frac{0.28}{1-3w_c} \left( \frac{10^{15}~{\rm g/cm}^3}{\rho_c} \right) 
\left( \frac{m}{10^{-11}~{\rm eV}} \right)^2\,,
\label{betabo}
\ee
where the right hand-side is equivalent to the critical value $\beta_c$ 
given in Eq.~(\ref{betac}).
For large values of $\rho_c$, $w_c$ can be 
close to the relativistic value $1/3$ or even larger, 
so we need to implement the pressure to derive the scalar-field profile correctly.
For $w_c<1/3$ the scalar field increases as a function of $\phi$ 
around $r=0$, so it is possible to reach the asymptotic 
value $\phi_{v}=\pi f_B$ at spatial infinity. 
Even if $\phi(r)$ decreases around $r=0$ for $w_c>1/3$, 
the decrease of the EOS parameter $w=P/\rho$ around 
the NS surface to the region $w<1/3$ allows a possibility for 
increasing $\phi(r)$ to reach $\phi_{v}=\pi f_B$ 
outside the star. 
We note that we have ignored the term $4V_c$ in Eq.~(\ref{phir=0}) 
relative to $\rho_c-3P_c$, but if $f_B$ is as close as the order $\Mpl$, 
there is the contribution of the potential to $m_{\rm eff}^2$ 
especially around $w_c \simeq 1/3$.

In the asymptotic region outside the NSs, the field $\phi$ should 
relax toward the value $\pi f_B$. 
In this regime, we can set $\rho=P=0$, $V \to 0$, 
and $h \to 1$ in Eq.~(\ref{phieqs}) and ignore the terms $\beta \phi'^2 r^2$ 
and $\beta \phi \phi' r$ relative to $\Mpl^2$. 
Keeping the term $V_{,\phi} \simeq m^2 (\phi-\pi f_B)$ around $\phi=\pi f_B$, 
the solution to Eq.~(\ref{phieqs}) is approximately given by 
Eq.~\eqref{phiso1}, but the coefficient $A$ is different
from that on weak gravitational backgrounds.

\subsection{Numerically constructed scalar-field profile}

To study the existence of the field profile connecting 
the solution (\ref{phiap}) to the other solution (\ref{phiso1}), 
we numerically integrate Eqs.~(\ref{mattereq})-(\ref{phieqs})
from the center of NSs to a sufficiently large distance. 
We exploit Eqs.~(\ref{fr=0})-(\ref{Pr=0}) as the boundary 
conditions around $r=0$. 
In Eq.~(\ref{fr=0}), we can set $f_c=1$ without loss of generality. 
The field value $\phi_c$ at the center of star is iteratively found 
from the demand of realizing the asymptotic value 
$\phi(r) \to \pi f_B$ with $\phi'(r) \to 0$ far outside the star.
For the perfect fluid inside the NS, we use the analytic 
representation of SLy EOS parametrized by 
\be
\xi=\log_{10} (\rho/{\rm g \cdot cm}^{-3})\,,\qquad 
\zeta=\log_{10} (P/{\rm dyn \cdot cm}^{-2})\,,
\label{xizeta}
\ee
where the explicit relation between $\xi$ and 
$\zeta$ is given in Ref.~\cite{Haensel:2004nu}. 
The outside of NS is assumed to be in a vacuum 
with a vanishing density and pressure. 
For the numerical purpose, we introduce 
the following constants
\be
\rho_0=m_n n_0=1.6749 \times 
10^{14}~{\rm g}/{\rm cm}^{3}\,,\qquad
r_0=\sqrt{\frac{8\pi \Mpl^2}{\rho_0}}=89.664~{\rm km}\,,
\ee
where $m_n=1.6749 \times 10^{-24}$~g is the neutron 
mass and $n_0=0.1~(\rm fm)^{-3}$ is the typical number 
density of NSs. The density $\rho$ and radius $r$ are 
normalized by $\rho_0$ and $r_0$, respectively.

\begin{figure}[ht]
\begin{center}
\includegraphics[height=3.3in,width=3.4in]{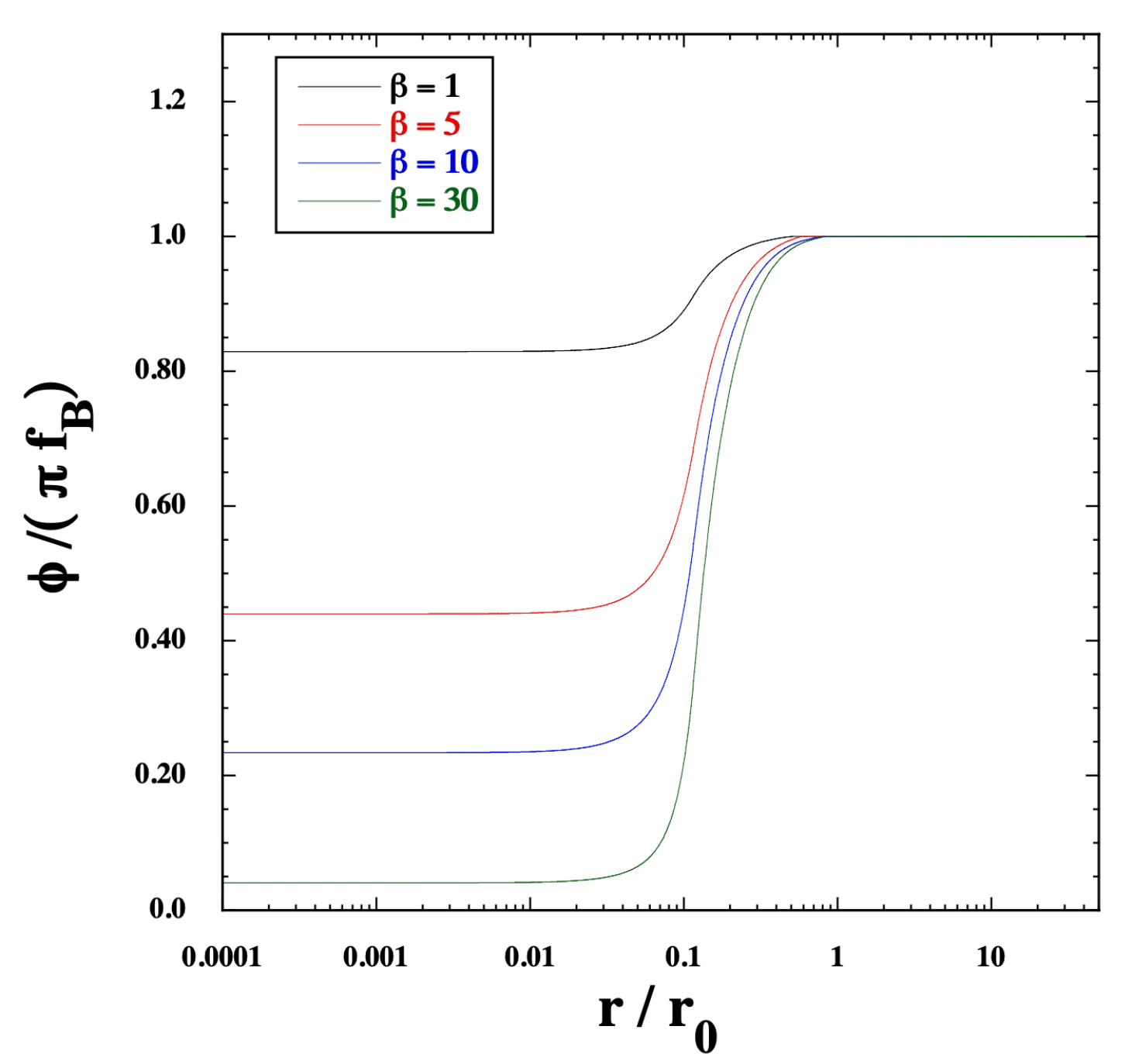}
\end{center}\vspace{-0.5cm}
\caption{\label{fig3}
The scalar field $\phi$ (normalized by $\pi f_B$) versus the radius $r$ 
(normalized by the distance $r_0=89.664$~km) for 
$m=1.0 \times 10^{-11}$~eV and $f_B=0.3 \Mpl/(\pi \beta)$. 
Each case corresponds to $\beta=1, 5, 10, 30$. 
We choose the SLy EOS with the central density 
$\rho_c=6 \rho_0$, where 
$\rho_0=1.6749 \times 10^{14}~{\rm g/cm}^3$.
}
\end{figure}

In Fig.~\ref{fig3}, we plot $\phi/(\pi f_B)$ versus $r/r_0$ for 
$\beta=1, 5, 10, 30$ with the model parameters 
$m=1.0 \times 10^{-11}$~eV and $f_B=0.3 \Mpl/(\pi \beta)$. 
The central density of NS is chosen to be 
$\rho_c=6 \rho_0 \simeq 10^{15}~{\rm g/cm}^3$.
With this mass scale $m$ the local gravity constraint (\ref{bebound}) 
gives the bound $\beta \pi f_B/\Mpl \le 0.3$ for the Sun, 
so the choice $f_B=0.3 \Mpl/(\pi \beta)$ corresponds to 
a maximally allowed value of $f_B$.
When $\rho_c=6 \rho_0$ the EOS parameter at $r=0$ is 
$w_c \simeq 0.158$, 
the condition (\ref{betabo})  
translates to $\beta>0.53$. In this case the effective 
mass squared $m_{\rm eff}^2$ is positive at $r=0$, 
and the scalar field grows according to Eq.~(\ref{phiap}) 
in the vicinity of $r=0$.

For $\beta=1$, the field value at $r=0$ 
is $\phi_c \simeq 0.83 \pi f_B$, 
and the difference from the asymptotic value is $\pi f_B-\phi_c \simeq 0.17 \pi f_B$.
On weak gravitational backgrounds discussed in Sec.~\ref{weaksec}, 
the field values inside and outside a star are very close to each other, 
see Eq.~(\ref{muphi0}) together with the condition $m^2 \gg \beta \rho/(2\Mpl^2)$.
Since the nonminimal coupling $\beta \rho/(2\Mpl^2)$ can be larger 
than $m^2$ for NSs around $r=0$, 
the difference between $\pi f_B$ and $\phi_c$ 
exceeds the order of $0.1 \pi f_B$. With the increase of $\beta$, 
this difference tends to be more significant, e.g., 
$\phi_c \simeq 0.04 \pi f_B$ for $\beta=30$. 
We note that the symmetry-breaking 
scale $f_B$ does not 
appear in the effective mass squared (\ref{meffs}) at $r=0$. 
Hence the normalized field configuration $\phi/(\pi f_B)$ 
is hardly sensitive to the change of $f_B$.

The large variation of $\phi(r)$ spanning in the range 
$0<\pi f_B-\phi_c \lesssim 0.1 \pi f_B$ is an outcome of the positive 
mass squared $m_{\rm eff}^2$ induced by 
large values of $\rho_c$. Then, the scalar field 
acquires a sufficient kinetic energy around $r=0$ 
to reach the asymptotic value $\phi_v=\pi f_B$ far outside 
the NSs. This is not the case for weak gravitational objects
where $\phi$ needs to stay around $\pi f_B$ both 
inside and outside the star. 
Thus, our model allows the existence of an interesting 
scalar-field profile whose variation is significant for 
strongly gravitating objects, 
while the variation of $\phi(r)$ is suppressed on weak 
gravitational backgrounds as consistent with Solar-system 
constraints.

\subsection{Modification of gravitational interactions}

The scalar-field profile derived above affects 
the nonlinearly extended gravitational 
potentials $\Psi$ and $\Phi$ through Eqs.~(\ref{feq}) and (\ref{heq}). 
Since $f={\rm e}^{2\Psi}$ and $h={\rm e}^{2\Phi}$, 
the left hand-side of Eq.~(\ref{heq}) is equivalent to 
$2(\Phi'-\Psi')$. In GR the right hand-side of Eq.~(\ref{heq}) 
vanishes for $r \geq r_s$, where $r_s$ is the 
radial position of the NS radius.
In the current model, however, there are contributions of 
$\phi(r)$ and its derivatives to the right hand-side 
of Eq.~(\ref{heq}). To quantify the difference between 
$\Phi'$ and $\Psi'$, we define 
\be
\eta (r) \equiv \frac{\Phi'(r)}{\Psi'(r)}-1\,,
\ee
and compute it at the surface of star.

\begin{figure}[ht]
\begin{center}
\includegraphics[height=3.2in,width=3.4in]{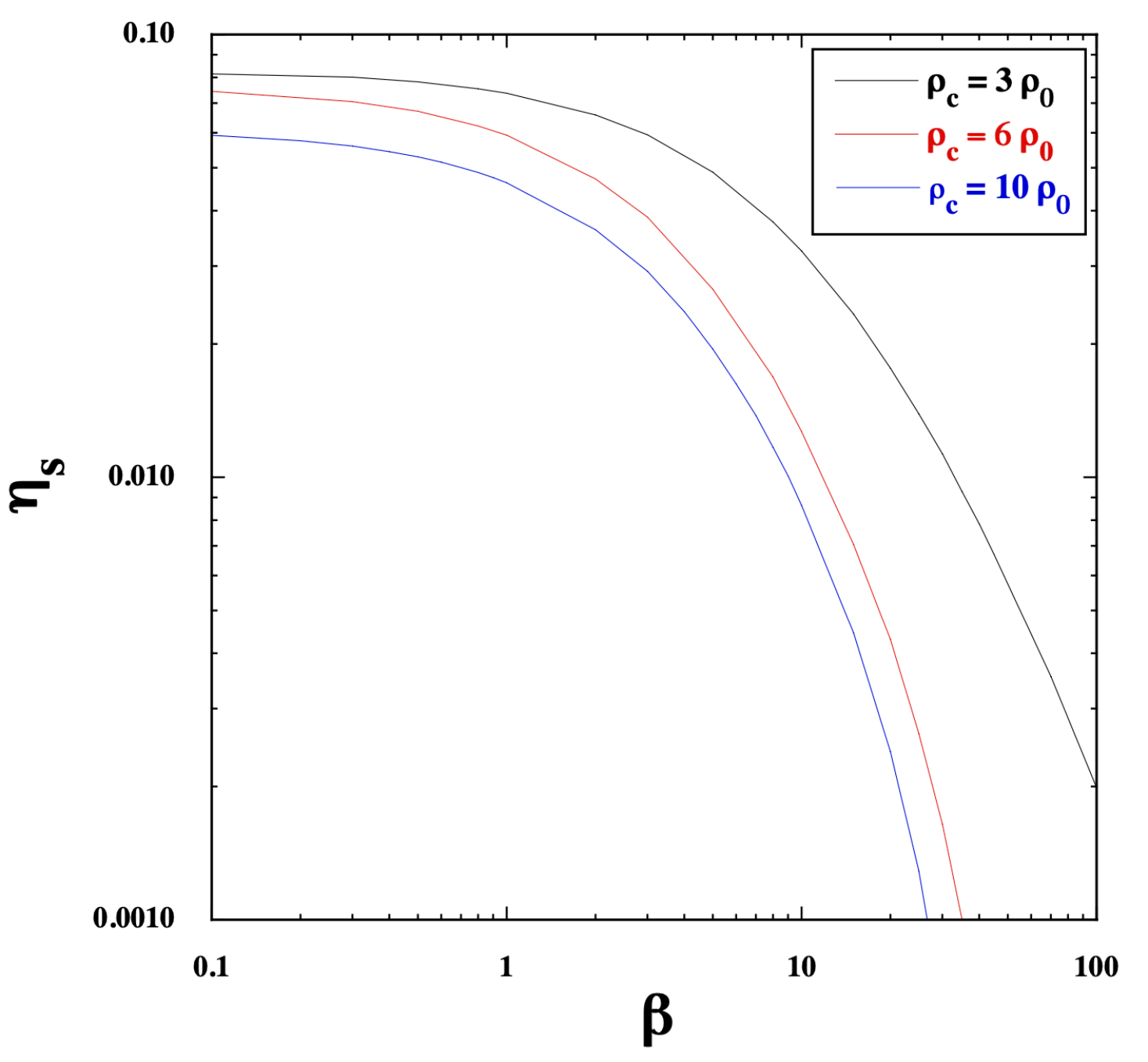}
\end{center}\vspace{-0.5cm}
\caption{\label{fig4}
We show $\eta_s=\Phi'(r_s)/\Psi'(r_s)-1$ versus the nonminimal 
coupling $\beta$ for $m=1.0 \times 10^{-11}$~eV and 
$f_B=0.3 \Mpl/(\pi \beta)$. We choose the SLy EOS with three 
different central densities  
$\rho_c=3 \rho_0, 6\rho_0, 10 \rho_0$.
}
\end{figure}

In Fig.~\ref{fig4}, we plot $\eta_s=\eta(r_s)$ versus 
$\beta$ for $\rho_c=3\rho_0, 6 \rho_0, 10 \rho_0$ 
with $m=1.0 \times 10^{-11}$~eV and 
$f_B=0.3 \Mpl/(\pi \beta)$. 
When $\rho_c=3 \rho_0$, the quantity $\eta_s$ can be 
as large as $0.08$ for $\beta$ in the range 0.1$\sim$1. 
Since we are choosing the maximally allowed value of $f_B$ 
consistent with Solar-system constraints, increasing $\beta$ 
results in smaller values of $f_B$. 
Given that $\phi(r)$ is normalized by $\pi f_B$
in the numerical simulation of Fig.~\ref{fig3}, 
decreasing $f_B$ implies smaller values of $\phi(r)$ inside the star.
Then, as $\beta$ increases, the corrections to gravitational 
potentials induced by the nonvanishing field 
profile should be more suppressed.
Indeed, for given $\rho_c$, the property of decreasing $\eta_s$ as a function of 
$\beta$ can be confirmed in Fig.~\ref{fig4}.

As $\rho_c$ increases in the range $\rho_c \geq 2\rho_0$, 
$\eta_s$ decreases from the maximum 
around $(\eta_s)_{\rm max}=0.08$ realized for 
the density $\rho_c= 2\rho_0 \sim 3\rho_0$. 
One of the reasons for this decrease of $\eta_s$ 
is that, for larger $\rho_c$, $w_c$ tends to increase.
For $\rho_c=3\rho_0, 6 \rho_0, 10 \rho_0$, we have 
$w_c=0.047, 0.158, 0.315$, respectively.
This means that, for $\rho_c \gtrsim 6 \rho_0$, 
the product $\rho_c(1-3w_c)$, which appears in 
the effective mass squared (\ref{meffs}), 
gets smaller as $\rho_c$ increases.
The other reason is that, as $\rho_c$ increases
in the range $2\rho_0 \leq \rho_c \lesssim 6 \rho_0$,
the field value $\phi_c$ at $r=0$ tends to be smaller 
by approaching the symmetry-restored state.
This leads to the overall decrease of corrections 
of the scalar field $\phi$ to the right hand-sides 
of Eqs.~(\ref{feq}) and (\ref{heq}).
Due to these two combined effects, for increasing $\rho_c$ 
in the range $2\rho_0 \leq \rho_c \lesssim 10 \rho_0$, 
there is the tendency that $\eta_s$ decreases.
Nevertheless, the observations of gravitational waves will provide us  
with interesting possibilities for probing the deviation from GR 
of order $\eta_s>{\cal O}(0.01)$ in the coupling range 
$\beta=0.1 \sim 10$.

As $\rho_c$ exceeds $10 \rho_0$, the product 
$\rho_c(1-3w_c)$ can be negative. 
In such cases the EOS parameter $w=P/\rho$ 
decreases toward the NS surface, and there is a point at which $\rho(1-3w)$ 
becomes positive. Then, it is possible to 
have nontrivial scalar-field profiles even for $\rho_c(1-3w_c)<0$, 
but the difference between $\pi f_B$ and $\phi_c$ tends to be smaller. 
For $\rho_c \gtrsim 10\rho_0$, this results in suppressed values 
of $\eta_s$ in comparison to the case $\rho_c \lesssim 10\rho_0$.

\begin{figure}[ht]
\begin{center}
\includegraphics[height=3.2in,width=3.4in]{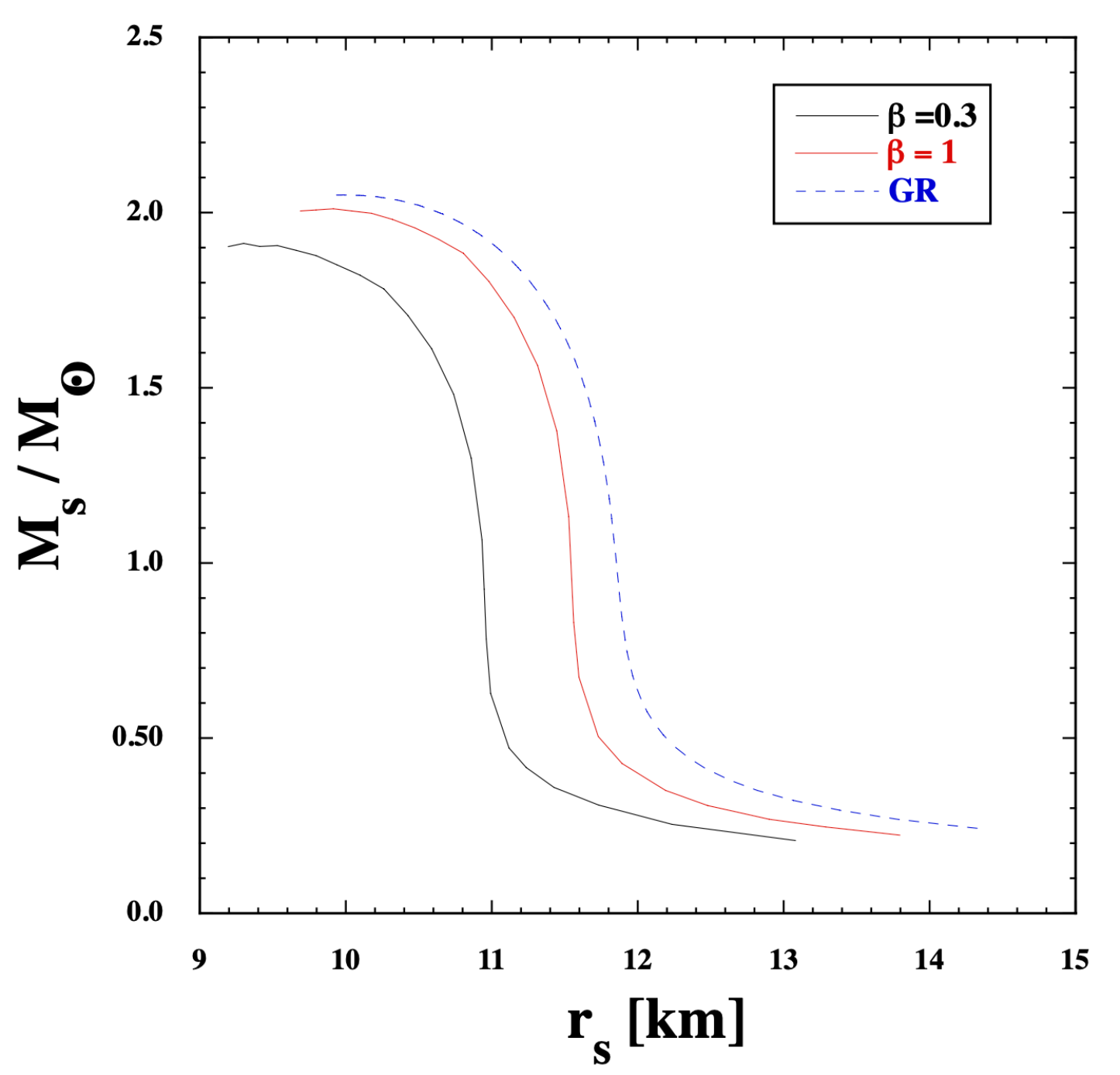}
\end{center}\vspace{-0.5cm}
\caption{\label{fig5}
The ADM mass $M_s$ normalized by the Solar-mass $M_{\odot}$ 
versus the NS radius $r_s$ for the SLy EOS with 
$m=1.0 \times 10^{-11}$~eV and $f_B=0.3 \Mpl/(\pi \beta)$. 
The solid black and red lines correspond to $\beta=0.3$ and 
$\beta=1$, respectively, while the dashed blue line is the case 
in GR. In all cases, the matter density at $r=0$ are in the range 
$2\rho_0 \leq \rho_c \leq 18 \rho_0$.
}
\label{fig5}
\end{figure}

The ADM mass of NSs is defined by 
\be
M_s \equiv 4\pi \Mpl^2 r \left[ 1-h(r) \right] |_{r=\infty}\,,
\ee
while the NS radius is determined by the condition 
$P(r_s)=0$. 
In Fig.~\ref{fig5}, we plot the relation between $M/M_{\odot}$ 
and $r_s$ for $\beta=0.3$ and 1 with $m=1.0 \times 10^{-11}$~eV 
and $f_B=0.3 \Mpl/(\pi \beta)$, where $M_{\odot}$ is 
the Solar mass. In comparison, we also show the case of GR 
derived for the SLy EOS without the scalar field $\phi$. 
The matter density $\rho_c$ at $r=0$ is chosen to be 
in the range $\rho_c \geq 2\rho_0$, under which 
the NS in GR has the ADM mass $M_s \geq 0.243M_{\odot}$ 
and radius $r_s \leq 14.33$~km.

In Fig.~\ref{fig5}, we observe that, for $\beta>0$, both $M_s$ and 
$r_s$ are smaller than those in GR. 
For $\beta={\cal O}(0.1)$ the field value $\phi_c$ is generally quite 
close to $\pi f_B$, so the exponential factor 
${\rm e}^{\beta \phi_c^2/(2\Mpl^2)}$ in Eq.~(\ref{Pr=0}) 
can be estimated as 
${\rm e}^{\beta \phi_c^2/(2\Mpl^2)} \simeq 1+0.045/\beta>1$.
This leads to the larger decreasing rate of $P(r)$ in comparison to GR, 
and hence $r_s$ and $M_s$ are reduced. 
Such reductions of $r_s$ and $M_s$ are different from 
the properties in standard spontaneous scalarization induced by 
the negative coupling $\beta$ \cite{Damour:1993hw,Minamitsuji:2016hkk,Doneva:2017duq}. 
For $\beta=0.3$ and $\rho_c=2\rho_0$ we obtain 
$r_s=13.08$~km and $M_s=0.207M_{\odot}$, so the relative 
difference from the ADM mass in GR 
($M_{\rm GR}=0.243M_{\odot}$) with the same value of 
$\rho_c$ is as large as 
$\mu_M =0.15$,~where we have defined
\be
\mu_M \equiv 1-\frac{M_s}{M_{\rm GR}}.
\ee
As $\rho_c$ increases, the deviation parameter $\mu_M$ 
tends to decrease, e.g., 
$\mu_M=0.10$ for $\rho_c=5 \rho_0$ and $\mu_M=0.07$ 
for $\rho_c=15 \rho_0$.
However, it is interesting to note that, for $\beta=0.3$, 
the difference of order 7\,\% from GR arises for the 
ADM mass even with high densities like $\rho_c \gtrsim 10\rho_0$.

With the increase of $\beta$, $\phi_c$ tends to decrease 
toward the symmetry-restored state $\phi=0$
and hence the exponential factor ${\rm e}^{\beta \phi_c^2/(2\Mpl^2)}$ 
approaches 1. Moreover, as we already discussed in Fig.~\ref{fig4}, 
the quantity $\eta_s$ is a decreasing function of $\beta$ 
for the choice $f_B=0.3 \Mpl/(\pi \beta)$. Indeed, as we see 
the case $\beta=1$ in Fig.~\ref{fig5}, the deviations of $M_s$ 
and $r_s$ from those in GR are less significant relative to 
the coupling $\beta=0.3$. 
Still, for $\beta=1$, we have $\mu_M=0.08$ for $\rho_c=2 \rho_0$ 
and $\mu_M=0.04$ for $\rho_c=5 \rho_0$, so the 
appreciable deviation from GR is present. 
For $\beta$ exceeding the order of 1, the theoretical curve in 
the $(M_s, r_s)$ plane approaches that of GR.

In standard spontaneous scalarization induced by negative 
values of $\beta$, the theoretical curve in the $(M_s, r_s)$ plane 
starts to be bifurcated from the GR one for $\rho_c$ larger 
than some critical density, and the mass of NSs in the scalarized branch 
are larger than that in GR with the same central density. 
This characterizes a continuous phase transition from the GR branch $\phi=0$ 
to the other nontrivial branch $\phi \neq 0$ triggered by a tachyonic instability. 
In our case, the scalarized NS solutions arise from the symmetry 
restored state at $\phi=0$ induced by positive $\beta$.
Theoretical plots for $\beta>0$ differ from the GR curve 
in the whole range of $\rho_c$ shown in Fig.~\ref{fig5}, i.e., 
$2\rho_0 \leq \rho_c \leq 18\rho_0$. 
This is because in our model there is 
no GR solution and hence no bifurcation from it.
Instead, we have only one branch of NS solutions where 
the scalar field $\phi$ asymptotically approaches 
the ground state located around $\phi=\pi f_B$.

In theories given by the action (\ref{action}), the conditions for 
the absence of ghost/Laplacian instabilities against 
odd- and even-parity perturbations are given by 
$\rho+P>0$ and $F(\phi)>0$ \cite{Kase:2021mix} (see also 
Ref.~\cite{Kase:2020qvz}). Since we are considering the 
nonminimal coupling $F(\phi)={\rm e}^{-\beta \phi^2/(2M_{\rm pl}^2)}$ 
with $\beta>0$ in the presence of a perfect-fluid matter satisfying the weak energy condition, 
there are neither ghost nor Laplacian instabilities for our 
scalarized NS solutions as in the case of standard 
spontaneous scalarization.

It should be noted that NSs in GR with other EOMs may provide the similar
ADM mass and radius to those derived for nonzero $\beta$ in Fig.~\ref{fig5}.
In our case, since the mass of NSs is relatively suppressed to that 
in the GR case, 
it would be difficult to distinguish NSs in our theory 
from those in GR with a different choice of EOSs,
only with observations regarding the mass and radius of NSs.
In order to break the degeneracy between the modified gravity effects 
and the ambiguity associated with the choice of EOSs,
we should explore the existence of universal relations
which are almost insensitive to the choice of EOSs \cite{Yagi:2013awa,Yagi:2014bxa,Breu:2016ufb}
as well as signatures associated with gravitational perturbations of NSs 
such as tidal deformability and quasinormal frequencies. 
We leave these subjects for future works.

Finally, we should comment on the case in which the energy density 
of an oscillating scalar field $\phi$ around $\phi=\pi f_B$
is responsible for a fraction of
CDM without decaying 
to other particles by today. 
In this case, the symmetry-breaking 
scale $f_B$ 
is constrained as Eq.~(\ref{fBcon}). 
When $\beta=1$, this gives the constraint 
$f_B/\Mpl \simeq 2 \times 10^{-6}$. 
For such small values of $f_B$, the field $\phi$ inside 
and outside the NS is also suppressed and hence $\eta_s$ is 
at most of order $10^{-9}$ with the nonminimal coupling 
in the range $\beta \le {\cal O}(10)$.
In such cases, the NS mass and radius are also very similar to those in GR.
However, there is a possibility that the oscillating field $\phi$ decays 
to other particles whose energy densities decrease as that of radiation or 
faster, in which case larger values of $f_B$ are allowed. 
In Figs.~\ref{fig4} and \ref{fig5}, we have used 
the maximum allowed values of $f_B$ 
consistent with Solar-system constraints.

\section{Conclusions}
\label{consec}

We proposed a new scenario of NS scalarizations 
in the presence of a pNGB potential 
$V(\phi)=m^2 f_B^2 [1+\cos(\phi/f_B)]$ and 
a nonminimal coupling to the Ricci scalar $R$
of the form $F(\phi)={\rm e}^{-\beta \phi^2/(2\Mpl^2)}$. 
In regions of the high density, the scalar field $\phi$ 
acquires a large positive mass squared by the 
nonminimal coupling with $\beta>0$. 
This can overwhelm a negative mass squared $-m^2$ of 
the bare potential at $\phi=0$.
Then, the symmetry restoration toward $\phi=0$ occurs  
in strong gravitational backgrounds like the interior 
of NSs, while the scalar approaches a VEV $\phi_v=\pi f_B$ 
toward spatial infinity.
This allows the existence of nontrivial 
field profiles affecting gravitational interactions 
in the vicinity of NSs. 

Unlike the original scenario of spontaneous scalarization induced 
by negative $\beta$ with $V(\phi)=0$ \cite{Damour:1993hw}, 
our model does not suffer from the tachyonic instability of 
cosmological solutions. 
In Sec.~\ref{cossec} we studied the cosmological evolution of $\phi$ 
for $m ={\cal O} ( 10^{-11}$~eV) and $f_B \lesssim \Mpl$ relevant to 
the mass scales of NS scalarizations. For $\beta>{\cal O}(0.1)$, 
the amplitude of $\phi$ exponentially decreases during 
inflation due to the dominance of the positive nonminimal 
coupling over the tachyonic mass squared $-m^2$ 
in the effective mass squared $m_{\rm eff}^2$.
During the reheating stage, the field amplitude exhibits mild 
decrease further. 
During the radiation-dominated era, 
after the contribution from the nonminimal coupling 
to $m_{\rm eff}^2$ drops below $-m^2$, $\phi$ starts to roll down 
the potential toward the VEV $\pi f_B$.
Numerically, we showed that the scalar field starts to oscillate 
around $\phi=\pi f_B$ before the epoch of BBN. 
If the oscillation of $\phi$ has continued by today, it can be 
the source of (a portion of) CDM for $f_B$ satisfying Eq.~(\ref{fBcon}). 
This relation is not applied to the case in which the energy density of 
oscillating $\phi$ is converted to other particles by today.

In Sec.~\ref{weaksec}, we derived the field profile for nonrelativistic 
stars with constant $\rho$ on weak gravitational backgrounds.
Outside the star, the scalar field is given by Eq.~(\ref{phica}) 
with the effective coupling (\ref{beff}). 
For the mass satisfying the condition $m r_s \gg 1$, where 
$r_s$ is the radius at the surface of star, 
the field stays in the region very close to 
$\phi=\pi f_B$.
In this case, the gravitational potentials $\Phi$ and $\Psi$ receive 
fifth-force corrections as Eqs.~(\ref{Phi}) 
and (\ref{Psi}). {}From Solar-system constraints on the 
post-Newtonian parameter $\gamma_{\rm PPN}=\Phi/\Psi$,  
we obtained the upper limit (\ref{bebound}) on the 
product $\beta f_B$.  
With the mass scale $m=10^{-11}$~eV, 
the bound (\ref{bebound}) translates to $\beta \pi f_B/\Mpl \leq 0.3$ 
for the Sun. This is weaker than the constraint (\ref{bebound2}) derived 
in the massless limit $m r_s \to 0$ by two orders 
of magnitude.

In Sec.~\ref{scasec}, we have numerically constructed NS solutions 
in the presence of a positive nonminimal coupling with the 
self-interacting potential.
We showed the existence of scalar-field profiles with significant difference between the field value $\phi_c$ at $r=0$ and the asymptotic value $\pi f_B$ 
at spatial infinity for static and spherically symmetric NSs. 
This is an outcome of the symmetry restoration toward 
$\phi=0$ in regions of the high density 
induced by the positive nonminimal coupling $\beta$.
As we observe in Fig.~\ref{fig3}, for larger $\beta$, the difference 
between $\phi_c$ and $\pi f_B$ tends to be more significant. 
The nonminimally coupled scalar field gives rise to 
modifications to the gravitational potentials $\Phi$ 
and $\Psi$ in comparison to GR.
We computed the quantity $\eta_s=\Phi'(r_s)/\Psi'(r_s)-1$
by varying the central density $\rho_c$ in the range $\rho_c \geq 2\rho_0$. 
Taking the upper limit $f_B=0.3 \Mpl/(\pi \beta)$ 
constrained from Solar-system tests of gravity, 
we find that $\eta_s$ is a decreasing function of $\beta$ 
for given $\rho_c$. The increase of $\beta$ results in 
the decreases of $f_B$ and $\phi(r)$ inside the star, 
so the parameter $\eta_s$ tends to be suppressed. 
As $\rho_c$ increases, $\eta_s$ is also subject to the decrease
due to several combined effects explained in the main text. 
Still, $\eta_s$ can be of order $\eta_s>{\cal O}(0.01)$ 
in the coupling range $\beta=0.1 \sim 10$.

In Fig.~\ref{fig5}, we plotted the relation between the ADM mass 
$M_s$ and the radius $r_s$ of NSs for $\beta=0.3$ and $\beta=1$ 
with $f_B=0.3 \Mpl/(\pi \beta)$.
Unlike standard spontaneous scalarization, the deviation of 
$M_s$ and $r_s$ from their values in GR occurs for 
any central density in the range $2 \rho_0 \leq \rho_c \leq 18 \rho_0$. 
For $\beta=0.3$, the relative difference of the ADM mass from that in GR, 
which is defined by 
$\mu_M = 1-M_s/M_{\rm GR}$, is as large as $\mu_M=0.15$ for 
$\rho_c=2\rho_0$. As $\rho_c$ increases, $\mu_M$ tends to decrease, 
but $\mu_M$ still has a considerably large value $0.07$ 
even for $\rho_c=15 \rho_0$. With the increase of $\beta$, 
the theoretical lines in the $(M_s, r_s)$ plane, 
which exist in the region $M_s<M_{\rm GR}$, approach that in GR.
For $\beta=1$ and $\rho_c=2\rho_0$, we found that 
$\mu_M=0.08$ and hence there is still appreciable deviation from GR.

In summary, we showed that our new model of scalarizations 
of NSs associated with the symmetry restoration 
induced by the nonminimal coupling
leads to modified gravitational interactions in the vicinity of NSs, 
while it is free from the problem of instabilities during 
the cosmological evolution. The implication of our model to observations 
of the binary NS coalescense was recently studied 
in Ref.~\cite{Higashino:2022izi}. Since the scalar field mass 
$m={\cal O} (10^{-11}\,{\rm eV})$ is 
larger than the typical orbital frequency of NS binaries 
$\omega={\cal O}(10^{-13}\,{\rm eV})$, the scalar radiation 
emitted from compact binaries during the inspiral phase 
is strongly suppressed. The resulting gravitational wave forms are 
similar to those in GR, so our model evades current observational 
constraints of inspiral gravitational wave forms.

\section*{ACKNOWLEDGMENTS}
MM was supported by the Portuguese national fund 
through the Funda\c{c}\~{a}o para a Ci\^encia e a Tecnologia (FCT) in the scope of the framework of the Decree-Law 57/2016 of August 29, changed by Law 57/2017 of July 19,
and the Centro de Astrof\'{\i}sica e Gravita\c c\~ao (CENTRA) through the Project~No.~UIDB/00099/2020.
MM also would like to thank Yukawa Institute for Theoretical Physics (under the Visitors Program of FY2022)
and Department of Physics of Waseda University for their hospitality.
ST was supported by the Grant-in-Aid for 
Scientific Research Fund of the JSPS Nos.~19K03854 and 22K03642.

\bibliographystyle{mybibstyle}
\bibliography{bib}

\end{document}